\documentclass[journal]{IEEEtran}
\usepackage{siunitx}
\usepackage{booktabs}
\usepackage{xcolor}
\usepackage{float}
\usepackage[normal,raggedright,scriptsize]{subfigure}
\usepackage[normalem]{ulem}

\usepackage{lineno,hyperref}

\usepackage{ifpdf}

\ifx\pdfoutput\undefined
\usepackage{graphicx}
\else
\usepackage[pdftex]{graphicx}

\usepackage{cite}

\ifCLASSINFOpdf
\else
\fi
\usepackage{algorithmic}

\usepackage{array}

\usepackage{stfloats}
\begin{document}
%
\title{Calculation of Mutual Inductance between Circular  and Arbitrarily Shaped Filaments via Segmentation Method }
%
%
%

\author{Kirill~V.~Poletkin,~Slobodan Babic,~Sreejith~Sasi~Kumar~
        and~Emil~R.~Mamleyev
\thanks{E. R. Mamleyev and K. V. Poletkin are with Institute of Microstructure Technology, Karlsruhe Institute of Technology, Hermann-von-Helmholtz-Platz 1, 76344 Eggenstein-Leopoldshafen, Germany.}
\thanks{K. V. Poletkin is also with New Uzbekistan University, Mustaqillik ave. 54, 100007 Tashkent, Uzbekistan  e-mail: k.poletkin@newuu.uz.}
\thanks{S. Babic is an independent researcher;  53 Berlioz 101, H3E 1N2, Montr\'{e}al, Qu\'{e}bec, Canada e-mail: slobobob@yahoo.com.}
\thanks{S. S. Kumar is with Technische Universit\"{a}t Ilmenau, Ehrenbergstraße 29, 98693 Ilmenau, Germany.}
}

\maketitle

\begin{abstract}
In this article, two analytical formulas for the calculation of mutual inductance between a circular filament and line segment arbitrarily positioning in the space are derived by using Mutual Inductance Method (MIM) and Babic's Method (BM), respectively. Using the fact that  any curve can be interpolated by a set of line segments, a method for calculation of mutual inductance between a circular filament and filament having an arbitrary shape in the space is proposed based on the derived analytical formulas.
The derived two formulas and the proposed method (Segmentation Method) were  numerically validated by using FastHenry software and reference examples from the literature. In particular, the proposed method was successfully applied to the calculation of mutual inductance between the circular filament and the following special curves such as  circle, circular arc,
elliptic arc, ellipse, spiral, helices and conical helices. All results of the calculation are in   good agreement with the reference examples.
\end{abstract}

\begin{IEEEkeywords}
mutual inductance, analytical formula, circular filament, special functions, complete elliptic functions, segment.
\end{IEEEkeywords}

%
\IEEEpeerreviewmaketitle

\section{Introduction}
%
%
%
%
%
%

\IEEEPARstart{A}{nalytical} and semi-analytical methods in   the calculation of self- and mutual-inductances of conducting elements of electrical circuits
have played an important role in development of power transfer, wireless communication,  sensing and actuation and have been applied in a broad fields of science, including electrical and electronic
engineering, medicine, physics, nuclear magnetic resonance, mechatronics and
robotics, to designate the most prominent.

Although, a number of efficient numerical methodes implemented in the commercially developed software are available,  analytical and semi-analytical methods allow to obtaining the result of calculations in the form of a
final formula with a finite number of input parameters, which when applicable
may significantly reduce computation effort. Providing the direct access to  a calculational formula for a user in such methods facilitates mathematical
analysis of obtained  results of calculation  and
opens an opportunity for their further  optimization.

In particular, calculation of mutual inductance between a circular filament and systems of filaments of different shapes is a prime example of application of such analytical methods.
Albeit, collections of formulas for the calculation of mutual inductance between filaments of different geometrical shapes covering a wide spectrum of practical arrangements have variously been presented in classical handbooks by Rosa\cite{Rosa1908}, Grover\cite{Grover2004}, Dwight\cite{Dwight1945}, Snow \cite{Snow1954}, Zeitlin\cite{Zeitlin1950}, Kalantarov \cite{Kalantarov1986}, among others.
However, these analytical methods have proved their efficiency and  have been successfully employed in an increasing number of applications, including electromagnetic levitation \cite{OkressWroughtonComenetzEtAl1952,Narukullapati2021}, superconducting levitation \cite{Paredes2021}, 
calculation of mutual inductance between  thick coils \cite{Ravaud2010,BabicSiroisAkyelEtAl2010,Poletkin2019,Wu2020,Yi2022,Pirincci2022}, {magnetic force and torque calculation between circular coils \cite{Babic2008,Babic2011,Babic2012,Babic2021,Poletkin2022}, 
calculation of magnetic stiffness \cite{Ravaud2010a,Poletkin2022a},
wireless power transfer \cite{JowGhovanloo2007,SuLiuHui2009,ChuAvestruz2017}, electromagnetic actuation \cite{ShiriShoulaie2009,RavaudLemarquandLemarquand2009,Obata2013}, 
micro-machined contactless inductive suspensions 
\cite{Poletkin2014a,Lu2014,PoletkinLuWallrabeEtAl2017b} 
and hybrid contactless suspensions \cite{Poletkin2012,PoletkinKorvink2018, Poletkin2020, Poletkin2021}, biomedical applications \cite{TheodoulidisDitchburn2007,SawanHashemiSehilEtAl2009}, topology optimization of coils \cite{KuznetsovGuest2017}, nuclear magnetic resonance \cite{D.I.B.2002,SpenglerWhileMeissnerEtAl2017}, indoor positioning systems \cite{AngelisPaskuAngelisEtAl2015}, navigation sensors \cite{WuJeonMoonEtAl2016},
wireless power transfer systems \cite{Zhang2021,Chu2021}, magneto-inductive wireless communications \cite{Gulbahar2017} and others.

Kalantarov and Zeitlin showed that the calculation of mutual inductance between a circular primary filament and any other secondary filament having an arbitrary shape and any desired position with respect to the primary filament can be reduced to a line integral \cite[Sec. 1-12, page 49]{Kalantarov1986}. Adapting this result, Poletkin  derived  the analytical formulas for calculating the mutual inductance between two circular filaments having any desired position with respect to each other in work  \cite{Poletkin2019},   as an alternative to Grover's  and Babi{c}’s expressions  reported in works \cite{Grover2004} and \cite{BabicSiroisAkyelEtAl2010}, respectively.

Moreover,  it was mentioned in work  \cite{Poletkin2019} that the obtained formula for  the treatment of   the singular case, when the circular filaments are mutually perpendicular, can be applied    also to the calculation of the mutual inductance between a circle and line segment after taking a  minor modification of the original formula.     Using this fact and an alternative approach developed by Babic  \cite{BabicSiroisAkyelEtAl2010}, two analytical formulas for calculation of mutual inductance between a circular filament and line segment arbitrarily positioning in the space are derived by means of   MIM and BM, respectively.

Since, any curve can be interpolated with a desired accuracy by a finite number of line segments, a segmentation method for calculation of the mutual inductance between the primary circle and a filament having an arbitrary shape is proposed  and successfully developed based on the two derived formulas for calculation of the mutual inductance between the circle and line segment.  It is shown that for calculation of mutual inductance  by means of the segmentation method,  a set of  points belonging to the  arbitrarily shaped filament as the input data is only needed. The developed methodology is successfully applied to the calculation of mutual inductance between the circular filament and the following special curves such as  circle, circular arc,
elliptic arc, ellipse, spiral, helices and conical helices. All results of calculation are in a  good agreement with the reference examples and successfully validated by   the \textit{FastHenry} software \cite{KamonTsukWhite1994} .


\begin{figure}[!t]
  \centering
  \includegraphics[width=2.2in]{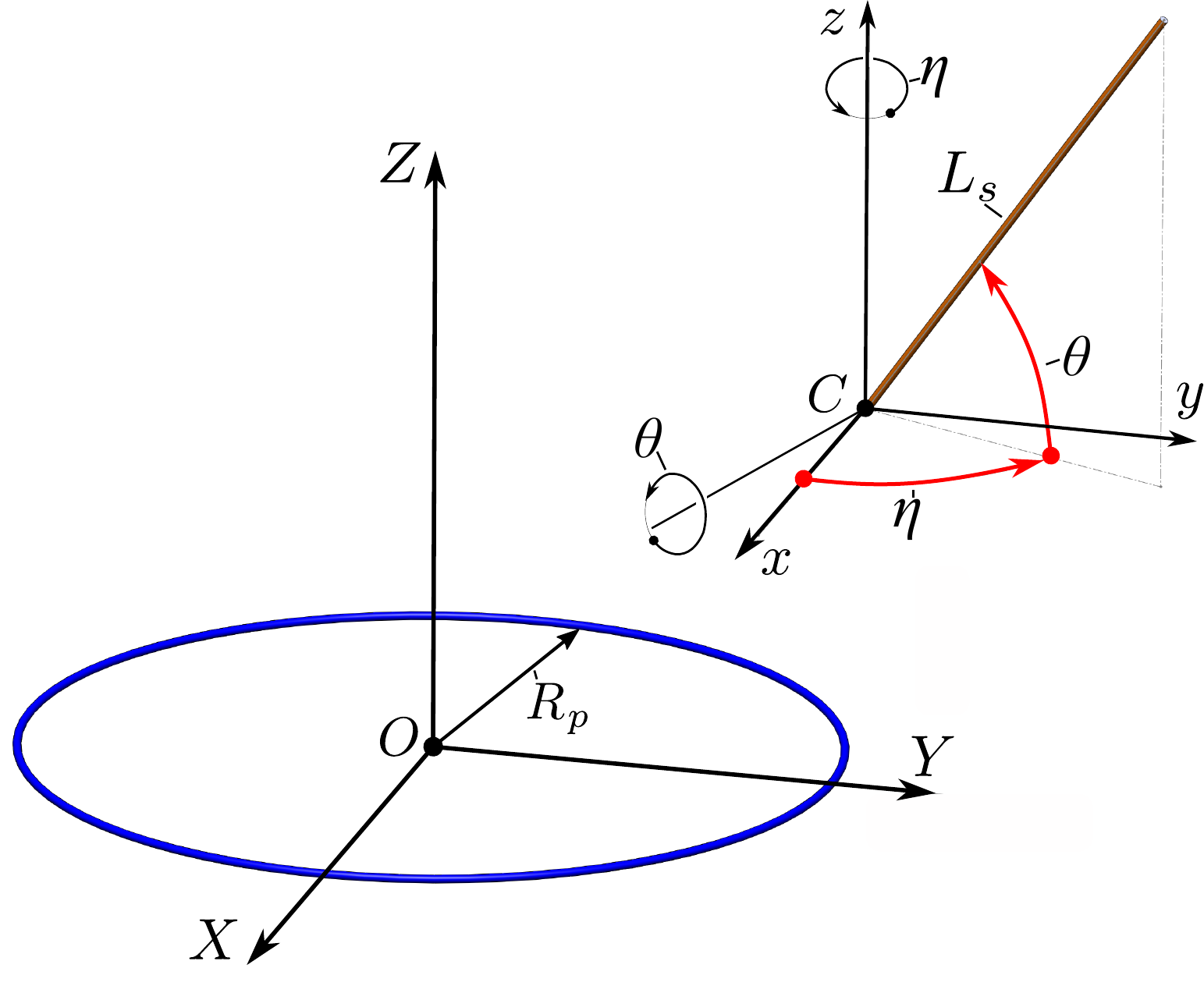}
  \caption{General scheme of arbitrarily positioning the line segment with respect to the circular filament: $R_p$ is the radius of the circle and $L_s$ is the length of the line segment.   }\label{fig:scheme}
\end{figure}

\section{Mutual Inductance between Circular Filament   and Line Segment}

Let us consider the filament system consisting of a primary circular filament having a radius of $R_p$ and a secondary filament represented as a line segment having a length of $L_s$, which is arbitrarily positioned in the space with respect to the primary circle as shown in Fig. \ref{fig:scheme}.  A coordinate frame (CF) denoted as $XYZ$ is assigned to the primary circle in a such way that the $Z$ axis is coincident with the circle axis and the $XOY$ plane of the CF lies on circle's plane, where the origin $O$ corresponds to the centre of primary circle. In turn, the $xyz$ CF is assigned to the secondary filament so that its origin $C$ is  attached to one of its endpoints. 
The axes of the $xyz$ CF are parallel to the axes of $XYZ$, respectively.

The linear position of the secondary filament with respect to the primary one is defined by the coordinates of the origin $C$ ($x_C,y_C,z_C$). The angular position of the line segment  is defined by the angle $\eta$  and $\theta$ corresponding to the angular rotation around the $z$-axis of the plane passing through the $z$-axis and line segment, and then the rotation of the segment around an axis, which is perpendicular to this segment-$z$ plane  and passing through the origin $C$, respectively, as shown in Fig. \ref{fig:scheme}.

\subsection{ Mutual Inductance Method (MIM)}
The mutual inductance between the circular filament and the line segment arbitrarily orientated in the space  can be calculated by
the following formula, which 
is derived by using Kalantarov-Zeitlin's approach reported in work \cite{Poletkin2019}.
Introducing the following dimensionless coordinates:
  \begin{equation}\label{eq:dimensionless_par}
    {\displaystyle {x}=\frac{x_C}{L_s},\; {y}=\frac{y_C}{L_s},\; {z}=\frac{z_C}{L_s}, {s}=\sqrt{{x}^2+{y}^2},} \\
  \end{equation}
the formula can be written as
\begin{equation}\label{eq:segment new}
 \begin{array}{l}
  {\displaystyle M=\frac{\mu_0\sqrt{R_pL_s}}{\pi}\int_{0}^{1}U\cdot\Phi(k)d\bar{\ell}},
  \end{array}
\end{equation}
where $\mu_0$ is the magnetic permeability of free space, $\bar{\ell}=\ell/L_s$ is the dimensionless integrating variable,
\begin{equation}\label{eq:U}
  U=U({x},{y},\eta,\theta)=\frac{t_1-t_2}{{\rho}^{1.5}}\cdot\cos\theta,
\end{equation}
\begin{equation}\label{eq:t for singular case}
  \begin{array}{l}
    t_1=t_1({x},\eta,\theta)=\sin\eta\cdot({x}+\bar{\ell}\cos\theta\cos\eta), \\
     t_2=t_2({y},\eta,\theta)=\cos\eta\cdot({y}+\bar{\ell}\cos\theta\sin\eta),\\
     {\rho}={\rho}({x},{y},\eta,\theta) =\\
     \sqrt{{s}^2+2\bar{\ell}\cos\theta\cdot  \left({x}\cos(\eta)+{y}\sin(\eta)\right)+\bar{\ell}^2\cos^2\theta},
  \end{array}
\end{equation}
\begin{equation}\label{eq:Phi}
   \Phi(k)=\frac{1}{k}\left[\left(1-\frac{k^2}{2}\right)K(k)-E(k)\right],
\end{equation}
and $K(k)$ and $E(k)$ are the complete elliptic functions of the first and second kind, respectively,
and
\begin{equation}\label{eq:k}
\begin{array}{l}
   {\displaystyle k^2=k^2({x},{y},{z},\theta,\eta)=\frac{4\nu{\rho}}{(\nu{\rho}+1)^2+\nu^2{z}_{\lambda}^2},
}\\
 {\displaystyle \nu=L_s/R_p,\;  {z}_{\lambda}={z}+\bar{\ell}\sin\theta}.
\end{array}
\end{equation}

\subsection{Babic's Method (BM)}
Let us consider again the calculation  scheme shown in Fig. \ref{fig:scheme}, but in a difference with MIM the line segment is given by coordinates of its two endpoints, namely, $p_0(x_0,y_0,z_0)$ and $p_1(x_1,y_1,z_1)$.

Then, the coordinates of an arbitrary point $S$ $(x_s, y_s, z_s)$ of the secondary line segment can be written parametrically  as follows:
 \begin{equation}\label{eq:coor S}
         x_s=q\bar{\ell}+ x_0, \;  y_s=r\bar{\ell}+ y_0,  \;  z_s=s\bar{\ell}+ z_0,  \;\bar{\ell}\in[0,1],
       \end{equation}
where $q=x_1-x_0$, $r=y_1-y_0$, $s=z_1-z_0$. Hence, the differential element of the line segment becomes
   \begin{equation}\label{eq:differential coord of SC}
         d\vec{l}_s=\{q, r, s\}d\bar{\ell}, \; \;\bar{\ell}\in[0,1].
       \end{equation}
While, parametric coordinates of an arbitrary point $P$ $(x_p, y_p, z_p)$ of the primary circular filament can be defined by the following expressions: 
       \begin{equation}\label{eq:parametric coord of PC}
         x_p=R_p\cos\phi, \;y_p=R_p\sin\phi, \;z_p=0,\; \phi\in[0,2\pi].
       \end{equation}
       The differential of the primary circular filament is given by
   \begin{equation}\label{eq:differential coord of PC}
         d\vec{l}_p=R_p\{-\sin\phi, \cos\phi, 0\}d\phi, \; \phi\in[0,2\pi].
       \end{equation}
Accounting for equations (\ref{eq:differential coord of PC}) and (\ref{eq:differential coord of SC}), the mutual inductance between circular filament and the line segment can be calculated by
\begin{equation}\label{eq:babic}
\begin{array}{c}
 {\displaystyle  M=\frac{\mu_{0}}{4 \pi} \int_{0}^{1} \int_{0}^{2\pi} \frac{d \vec{l}_{p} \cdot d \vec{l}_{s}}{r_{ps}}=}\\
 {\displaystyle \frac{\mu_{0}R_p}{4 \pi} \int_{0}^{1} \int_{0}^{2\pi} \frac{-q\sin\phi+r\cos\phi}{r_{ps}} d\phi d\bar{\ell}},
\end{array}
\end{equation}
where
\begin{equation}\label{eq:rps}
 r_{ps}^{2}=\left(x_{s}-R_{p} \cos (\phi)\right)^{2}+\left(y_{s}-R_{p} \sin (\phi)\right)^{2}+z_{s}^{2}.
\end{equation}
The following integral
\begin{equation}\label{eq:I}
\begin{array}{c}
 {\displaystyle I= \int_{0}^{2\pi} \frac{-q\sin\phi+r\cos\phi}{r_{ps}} d\phi }
\end{array}
\end{equation}
can be expressed via  the complete elliptic functions of the first and second kind $K(k)$ and $E(k)$, respectively, as follows
\begin{equation}\label{eq:I elliptic}
\begin{array}{c}
 {\displaystyle I=\frac{2V}{\sqrt{R_p}\sqrt{p^3}}\Psi(k), }
\end{array}
\end{equation}
where 
\begin{equation}\label{eq:I parameters}
\begin{array}{l}
   {\displaystyle V=r\cdot x_s-q\cdot y_s,\;  p=\sqrt{x_s^2+y_s^2},
}\\
   {\displaystyle  \Psi(k)=\left(\frac{2}{k}-k\right)K(k)-\frac{2}{k}E(k),
}\\
   {\displaystyle k^2=\frac{4R_pp}{(R_p+p)^2+{z}_{s}^2}.
}
\end{array}
\end{equation}
Hence, replacing the integral (\ref{eq:I}) in 
(\ref{eq:babic}) by 
(\ref{eq:I elliptic}),  the final formula for calculation of mutual inductance becomes
\begin{equation}\label{eq:babic final}
\begin{array}{c}
 {\displaystyle  M=\frac{\mu_{0}\sqrt{R_p}}{2 \pi} \int_{0}^{1}\frac{V}{\sqrt{p^3}}\Psi(k) d\bar{\ell}}.\\
\end{array}
\end{equation}

The input parameters for Babic's and MI method are related by the following relationships: 
\begin{equation}\label{eq:relationships between BM and MIM}
\begin{array}{l}
 {\displaystyle L^2_s=q^2+r^2+s^2},\\
 {\displaystyle \eta=\tan^{-1}\left(\frac{r}{q} \right)},\;
  {\displaystyle  \theta= \tan^{-1}\left(\frac{s}{\sqrt{q^2+r^2}} \right)}.
\end{array}
\end{equation}

Worth noting that according to  (\ref{eq:segment new}) and (\ref{eq:babic final}) the mutual inductance is equal to zero, when the line segment crosses the $Z$ axis or is located on a line crossing  the $Z$ axis.

\begin{figure}[!t]
  \centering
  \includegraphics[width=2.0in]{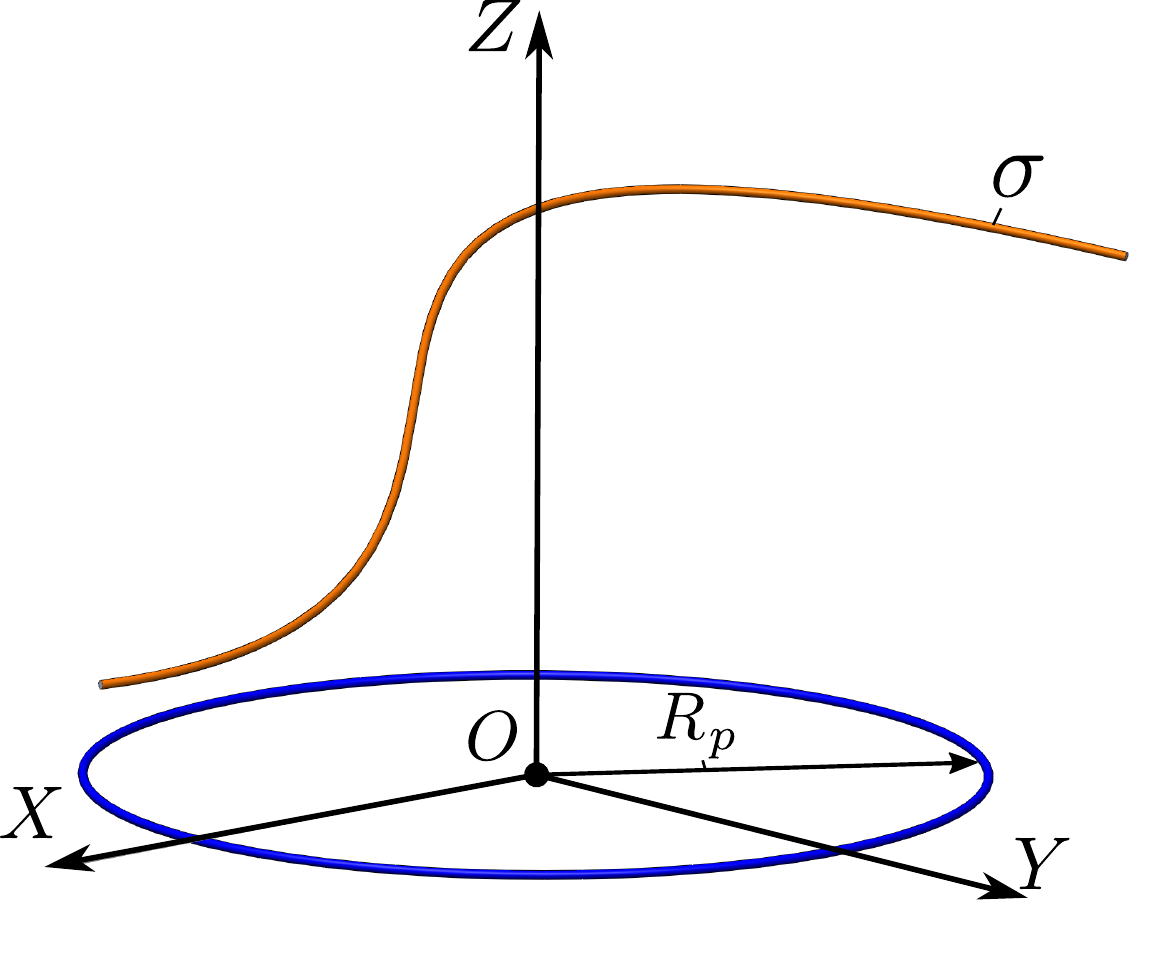}
  \caption{General scheme of arbitrary shaped filament and the circular filament.   }\label{fig:scheme arb}
\end{figure}
\section{ Segmentation Method }

 The secondary filament having an arbitrary shape and the primary circular filament of a radius $R_p$ are considered as shown in Fig. \ref{fig:scheme   arb}. The secondary filament is given, for instance, by a 3D parametric curve $\sigma=\sigma(\ell)$, where $\ell$  is the parameter defined within a finite interval $\left[\varphi_0,\varphi_1\right]$, and the following inequality $\varphi_1>\varphi_0\geq0$ is valid.

\begin{figure}[!t]
  \centering
  \includegraphics[width=2.0in]{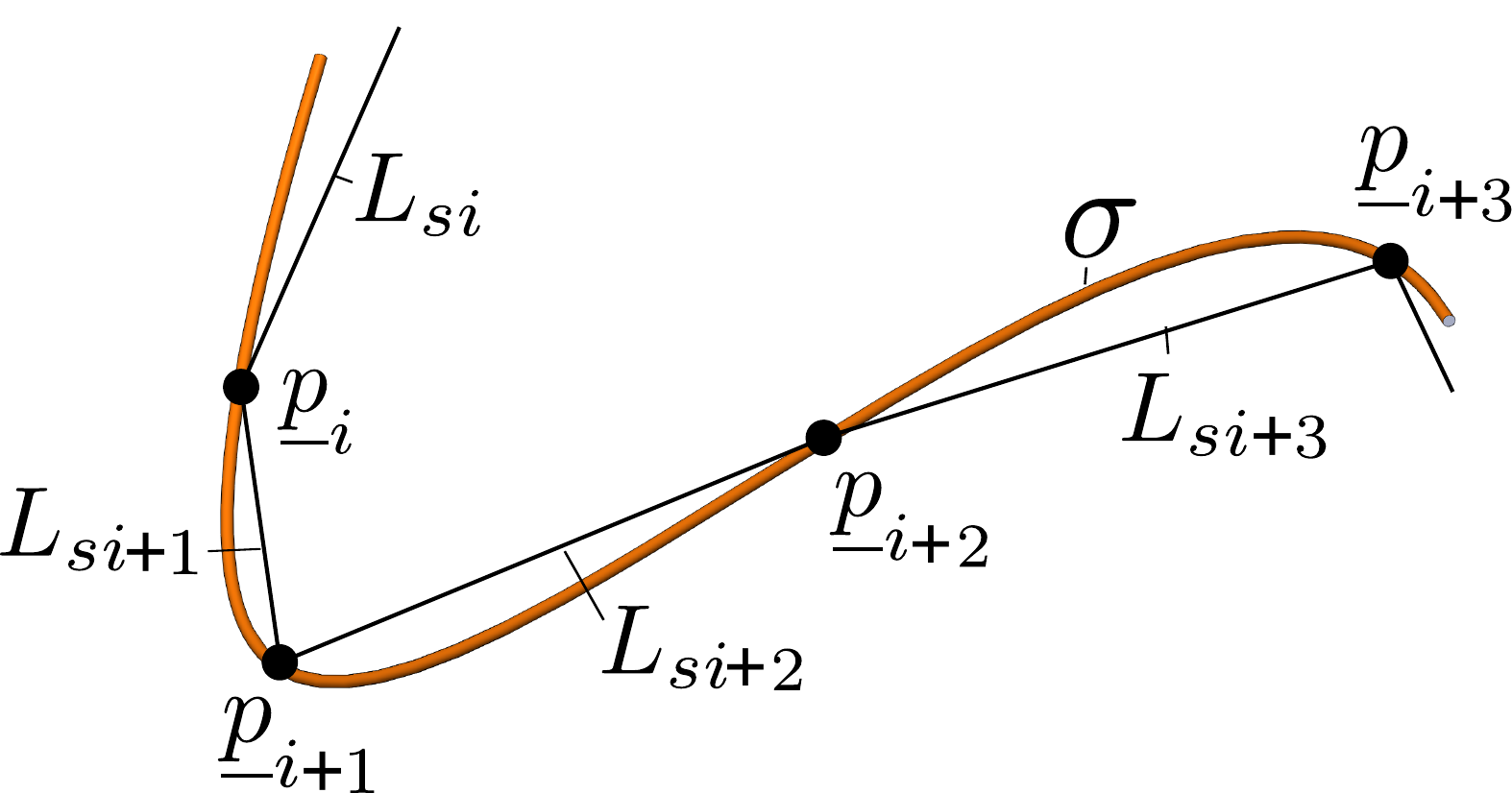}
  \caption{Interpolation of the $\sigma$ curve by line segments.   }\label{fig:interpolation}
\end{figure}
  It is assumed that the curve is sampled by $n$ points, so that we have  $\underline{p}_0(h_0)$, $\underline{p}_1(h_1)$, $\underline{p}_2(h_2)$, $\ldots$  , $\underline{p}_{n-1}(h_{n-1})$, $\underline{p}_n(h_n)$. The points can be defined in the following way
 \begin{equation}\label{eq:p}
   \underline{p}_i(h_i)=\left[x_{\sigma}(h_i)\; y_{\sigma}(h_i) \; z_{\sigma}(h_i)\right]^T, \; i=0\ldots n,
 \end{equation}
as  ($3\times1$)-column-matrices,  where $x_{\sigma}$, $y_{\sigma}$ and $z_{\sigma}$ are the coordinates of the $\sigma$-curve and
\begin{equation}\label{eq:h}
  h_{i}=h_{i-1}+(\varphi_1-\varphi_0)/n, \; i=1\ldots n,
\end{equation}
and so $h_0=\varphi_0$. Using the set of points  (\ref{eq:p}) and (\ref{eq:h}),  the $\sigma$-curve can be interpolated by line segments as shown in  Fig. \ref{fig:interpolation}. The length of each line segment can be calculated by
\begin{equation}\label{eq:Li}
  L_{si}^2=(\underline{p}_i-\underline{p}_{i-1})^T(\underline{p}_i-\underline{p}_{i-1}),
\end{equation}
while the angles and dimensionless coordinates  of $i$-th line segment become
\begin{equation}\label{eq:angles of seg}
\begin{array}{c}
 {\displaystyle \eta_i=\tan^{-1}\left(\frac{\underline{p}_{yi}-\underline{p}_{yi-1}}{\underline{p}_{xi}-\underline{p}_{xi-1}} \right),}\\
{\displaystyle \theta_i= \tan^{-1}\left(\frac{\underline{p}_{zi}-\underline{p}_{zi-1}}{\sqrt{\left(\underline{p}_{xi}-\underline{p}_{xi-1}\right)^2+\left(\underline{p}_{yi}-\underline{p}_{yi-1}\right)^2}} \right)};
\end{array}
\end{equation}
and
  \begin{equation}\label{eq:dimensionless_coor}
    {\displaystyle {x_i}=\frac{\underline{p}_{xi-1}}{L_{si}},\; {y_i}=\frac{\underline{p}_{yi-1}}{L_{si}},\; {z_i}=\frac{\underline{p}_{zi-1}}{L_{si}}, {s_i}=\sqrt{{x_i}^2+{y_i}^2},} \\
  \end{equation}
respectively. Substituting 
(\ref{eq:Li}), (\ref{eq:angles of seg}) and (\ref{eq:dimensionless_coor}) into  (\ref{eq:segment new}), the mutual inductance $M_i$ between the circular filament and $i$-th line segment can be calculated. Hence, performing summation of all $n$ terms of  $M_i$,  the formula for calculation of the mutual inductance between  the arbitrary shape and the primary circular filament can be written as follows
\begin{equation}\label{eq:segmentation method}
 \begin{array}{l}
  {\displaystyle M_{\sigma}=\frac{\mu_0\sqrt{R_p}}{\pi}\int_{0}^{1}\sum_{i=1}^{n}\sqrt{L_{si}}\cdot U_i\cdot\Phi(k_i)d\bar{\ell}}.
  \end{array}
\end{equation}
Alternatively, the calculation can be performed by using Babic's formula (\ref{eq:babic final}), namely,
\begin{equation}\label{eq:segmentation method babic}
 \begin{array}{l}
  {\displaystyle M_{\sigma}=\frac{\mu_0\sqrt{R_p}}{2\pi}\int_{0}^{1}\sum_{i=1}^{n}\frac{V_i}{\sqrt{p_i^3}}\Psi(k_i)d\bar{\ell}}.
  \end{array}
\end{equation}

Thus, interpolating the parametric $\sigma$-curve by $n$ line segments, the calculation of  the mutual inductance between the arbitrary shape and   the primary circular filament can be performed by means of using developed formulas, namely, (\ref{eq:segment new}) and (\ref{eq:babic final}), which are applied  $n$ times to each interpolating line segment followed by summation executed by 
(\ref{eq:segmentation method}) and (\ref{eq:segmentation method babic})  for all obtained calculation results. The \textit{Matlab} files with the implemented formulas (\ref{eq:segment new}), (\ref{eq:babic final}), (\ref{eq:segmentation method}) and (\ref{eq:segmentation method babic}) are available from the authors  as   supplementary materials to this article.

\begin{figure}[!t]
  \centering
  \includegraphics[width=3.0in]{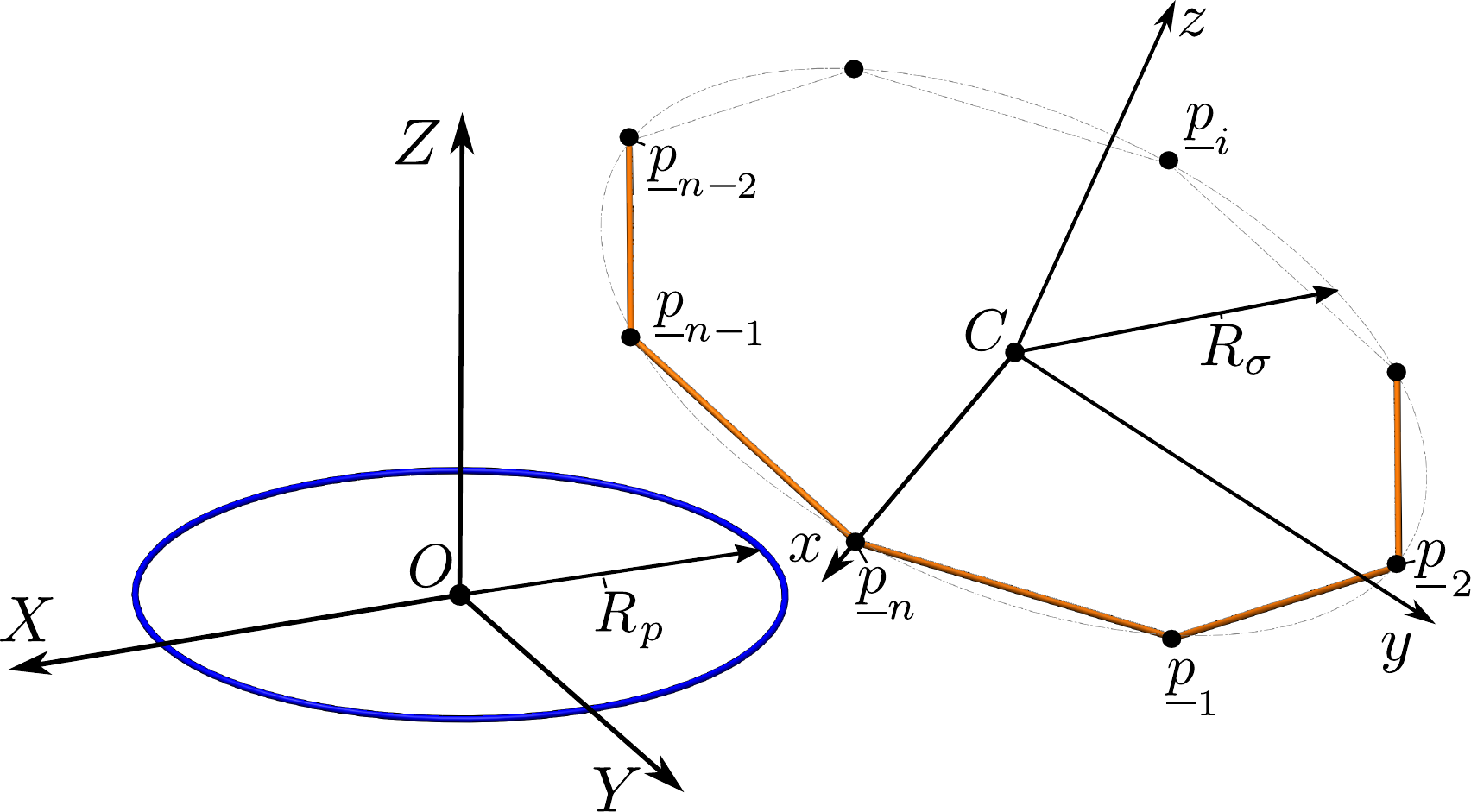}
  \caption{A regular polygon of $n$ sides arbitrarily positioned in the space with respect to the circle of radius $R_p$.   }\label{fig:polygon}
\end{figure}

\section{ Examples of Calculation. Numerical Verification }
In this section the developed formulas (\ref{eq:segment new}) and (\ref{eq:babic final}), as well as the proposed segmentation method based on 
(\ref{eq:segmentation method})  and  (\ref{eq:segmentation method babic})  are validated by using FastHenry software  and reference  examples from the literature.

\subsection{ Mutual inductance of line segment and circle}
\label{sec:segment}

\subsubsection*{ Example 1 }
Let us suppose that
the radius of primary circle is $R_p$=\SI{1}{\meter} and the line segment defined by two endpoints having the following coordinates, namely, $p_0(\SI{1}{\meter},\SI{2}{\meter},\SI{3}{\meter})$ and $p_1(\SI{2}{\meter},\SI{3}{\meter},\SI{4}{\meter})$. The results of calculation are

\vspace*{1.0em}
\begin{tabular}{cccc}
  \toprule
  &FastHenry& BM, (\ref{eq:babic final})  & MIM, (\ref{eq:segment new})\\
   \midrule
   $M$, nH&   $-3.53653$ & $-3.401894$ & $-3.401894$\\
  \toprule\label{tab:example1}
\end{tabular}

\subsubsection*{ Example 2 }
Let us suppose that
the radius of primary circle is $R_p$=\SI{1}{\meter} and the line segment defined by two endpoints having the following coordinates, namely, $p_0(\SI{1}{\meter},\SI{1}{\meter},\SI{1}{\meter})$ and $p_1(\SI{0}{\meter},\SI{1}{\meter},\SI{1}{\meter})$. The segment  is parallel to the plane of the circle. The results of calculation are

\vspace*{1.0em}
\begin{tabular}{cccc}
  \toprule
&FastHenry& BM, (\ref{eq:babic final})  & MIM, (\ref{eq:segment new})\\
   \midrule
   $M$, nH&   $69.4492$ & $69.51806$ & $69.51806$\\
  \toprule\label{tab:example2}
\end{tabular}

\subsubsection*{ Example 3 }
Let us suppose that
the radius of primary circle is $R_p$=\SI{0.03}{\meter} and the line segment defined by two endpoints having the following coordinates, namely, $p_0(\SI{0.0175}{\meter},\SI{-0.0029904}{\meter},\SI{0.0040192}{\meter})$ and $p_1(\SI{0.0025}{\meter},\SI{0.02299}{\meter},\SI{0.055981}{\meter})$. The results of calculation are

\vspace*{1.0em}
\begin{tabular}{cccc}
  \toprule
  &FastHenry& BM, (\ref{eq:babic final})  & MIM, (\ref{eq:segment new})\\
   \midrule
   $M$, nH&   $1.82457$ & $1.83574$ & $1.83574$\\
  \toprule\label{tab:example3}
\end{tabular}

\begin{figure}[!t]
  \centering
  \includegraphics[width=1.8in]{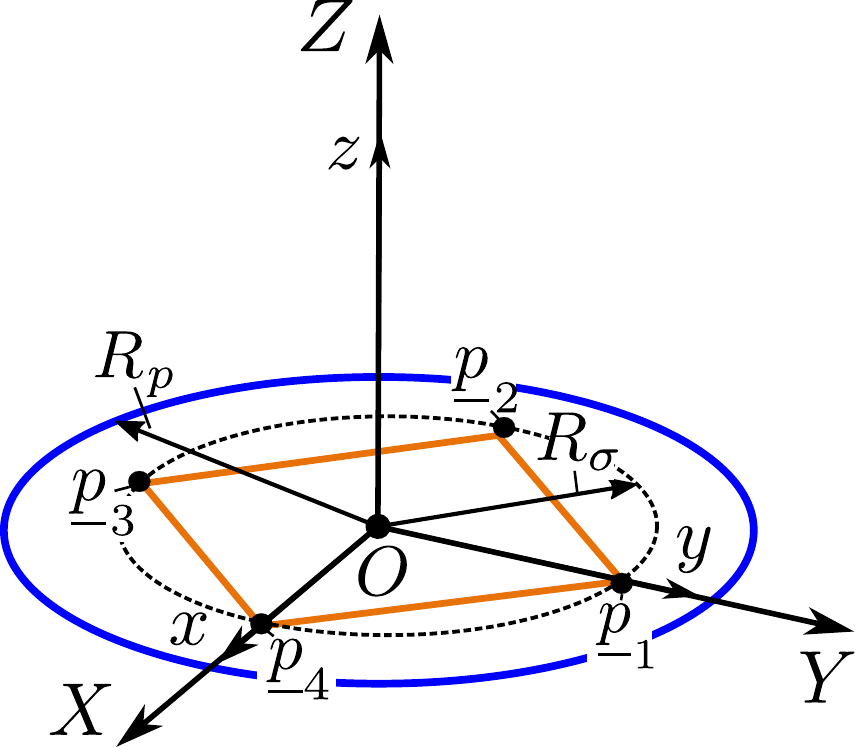}
  \caption{Scheme for Example 4.   }\label{fig:example4}
\end{figure}
\subsection{ Mutual inductance between circle and a regular polygon of $n$ sides arbitrarily positioned in  the space}
\label{sec:polygon}

In this section, the mutual inductance between a regular polygon of $n$ sides, which is arbitrarily positioned in the space with respect to the circle of radius $R_p$ is calculated by segmentation method based on 
(\ref{eq:segmentation method})  and  (\ref{eq:segmentation method babic}). It is assumed that the polygon is inscribed in a circle of a radius $R_{\sigma}$ lying on the $xy$ plane as shown in Fig. \ref{fig:polygon}. The parameter $\ell$ is defined within the finite interval $\left[0,2\pi\right]$. This interval determines the distribution of the polygon vertices on the $xy$ plane. In particular,  $n$-th vertex within such interval of the parameter $\ell$ is located on the $x$ axis.      The linear misalignment of the polygon is defined by coordinates of point $C$ corresponding to the centre of the secondary circle, while the angular misalignment can be defined by Grover's angles denoted by  $\theta$ and $\eta$ \cite{Poletkin2019}.

Then, input points can be calculated by
\begin{equation}\label{eq:points polygon}
 \underline{p}_i(h_i)=\left[\begin{array}{c} x_C \\ y_C \\  z_C \end{array}\right]+ \underline{\Lambda}_{\theta}^T \underline{\Lambda}_{\eta}^T\left[\begin{array}{c} R_{\sigma}\cos h_i \\ R_{\sigma}\sin h_i \\  0\end{array}\right], \; i=0\ldots n,
\end{equation}
where
\begin{equation}\label{eq:cosine matrix eta and theta}
  \underline{\Lambda}_{\eta}=\left[\begin{array}{ccc} \cos\eta & \sin\eta & 0\\ -\sin\eta & \cos\eta & 0\\ 0 & 0 & 1 \end{array}\right], \;
 \underline{\Lambda}_{\theta}=\left[\begin{array}{ccc} 1 & 0 & 0\\ 0 & \cos\theta & \sin\theta\\ 0 & -\sin\theta & \cos\theta \end{array}\right].
\end{equation}

\subsubsection*{ Example 4 (Example 1-3, page 50 in Kalantarov's book  \cite{Kalantarov1986})}
The primary circle of radius $R_p$=\SI{1}{\meter} and the square filament of a side length of \SI{1}{\meter}  (the polygon of $n=4$ sides, $R_{\sigma}=\sqrt{2}\cdot$\SI{0.5}{\meter}) are considered. The square is lying on the $XY$ plane and its origin coincides with point $O$ as shown in Fig. \ref{fig:example4}. The results of calculation
show

\vspace*{1.0em}
\begin{tabular}{lcccc}
  \toprule
 &Kalantarov's &FastHenry& BM,  & MIM, \\
  & book &      &  (\ref{eq:segmentation method babic}) &  (\ref{eq:segmentation method})\\
   \midrule
   $M_{\sigma}$, $\mu$H&$ 0.880$ &   $0.72949$ & $0.73075$ & $0.73075$\\
  \toprule\label{tab:example4}
\end{tabular}
\subsubsection*{ Example 5 }
Considering the same arrangement as in Example 4, but the centre of square filament is located at point $C$ having the following coordinates $x_C=y_C=$\SI{0}{\meter} and $z_C=$\SI{1}{\meter}$/\sqrt{2}$. The results of calculation
are

\vspace*{1.0em}
\begin{tabular}{lccc}
  \toprule
  &FastHenry& BM, (\ref{eq:segmentation method babic})  & MIM, (\ref{eq:segmentation method}) \\
   \midrule
   $M_{\sigma}$, $\mu$H &   $0.317264$ & $0.31754544$& $0.31754544$\\
  \toprule\label{tab:example1}
\end{tabular}

\subsubsection*{ Example 6 }
The primary circle has a radius of $R_p=$\SI{16}{\centi\meter}.  The triangular filament is inscribed in a circle having a radius of   $R_{\sigma}=$\SI{10}{\centi\meter}. The centre of the secondary circle has the following coordinates:  $x_C=0$, $y_C=$\SI{4.3301}{\centi\meter}, and $z_C=$\SI{17.5}{\centi\meter}. While,  the angular misalignment is defined by the following Grover's angles, namely,    $\eta=$\SI{45}{\degree} and $\theta=$\SI{60.0}{\degree} as shown in Fig. \ref{fig:example6}.
 The results of calculation are

\vspace*{1.0em}
\begin{tabular}{lccc}
  \toprule
  &FastHenry& BM, (\ref{eq:segmentation method babic})  & MIM, (\ref{eq:segmentation method}) \\
   \midrule
   $M_{\sigma}$, nH &   $5.94445$ & $5.927887$& $5.927887$\\
  \toprule\label{tab:example1}
\end{tabular}

\begin{figure}[!t]
  \centering
  \includegraphics[width=1.4in]{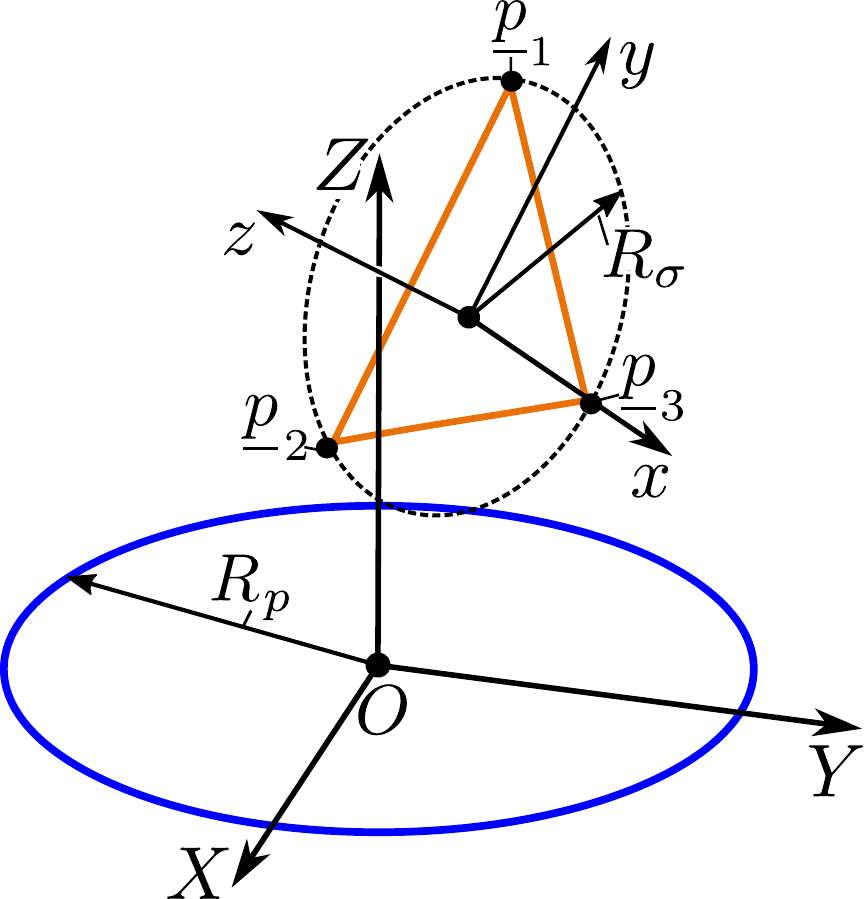}
  \caption{Scheme for Example 6.   }\label{fig:example6}
\end{figure}
\subsubsection*{ Example 7 }
Using the same arrangement as in previous Example 6,  the square polygon is inscribed.
The results of calculation are

\vspace*{1.0em}
\begin{tabular}{lccc}
  \toprule
  &FastHenry& BM, (\ref{eq:segmentation method babic})  & MIM, (\ref{eq:segmentation method}) \\
   \midrule
   $M_{\sigma}$, nH &   $9.40649$ & $9.435092$& $9.435092$\\
  \toprule\label{tab:example7}
\end{tabular}
\subsubsection*{ Example 8 }
The hexagon ($n=6$ sides) is inscribed in the secondary circle of the same arrangement as in  Example 6.
The results of calculation are

\vspace*{1.0em}
\begin{tabular}{lccc}
  \toprule
  &FastHenry& BM, (\ref{eq:segmentation method babic})  & MIM, (\ref{eq:segmentation method}) \\
   \midrule
   $M_{\sigma}$, nH &  $12.4966$ & $12.54164$& $12.54164$\\
  \toprule\label{tab:example8}
\end{tabular}

\begin{figure}[!t]
  \centering
  \includegraphics[width=1.9in]{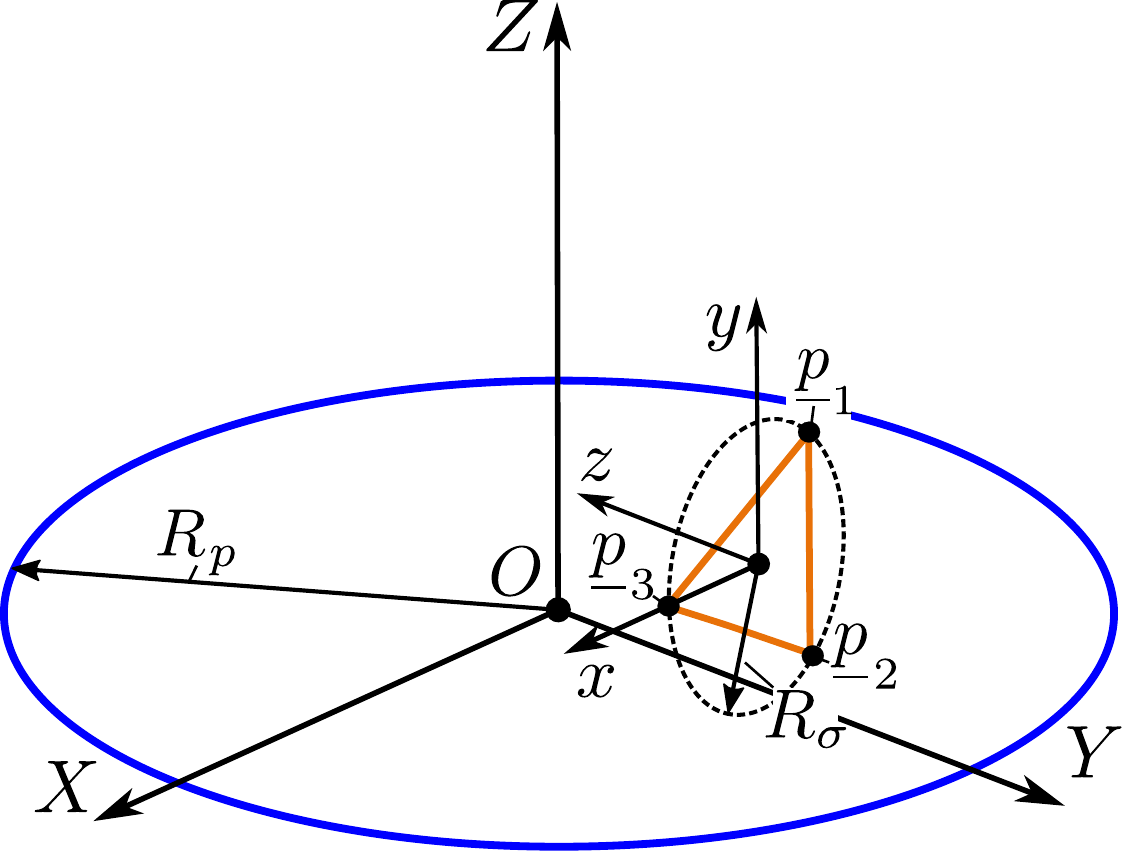}
  \caption{Scheme for Example 9.   }\label{fig:example9}
\end{figure}
\subsubsection*{ Example 9 }
The primary circle has a radius of $R_p=$\SI{40}{\centi\meter}.  The triangular filament is inscribed in a circle having a radius of   $R_{\sigma}=$\SI{10}{\centi\meter}, which is perpendicular to the primary circle that Grover's angles of    $\eta=$\SI{0}{\degree} and $\theta=$\SI{90.0}{\degree}.  The centre of the secondary circle has the following coordinates:  $x_C=0$, $y_C=$\SI{20}{\centi\meter}, and $z_C=$\SI{10}{\centi\meter} as shown Fig. \ref{fig:example9}. 
The results of calculation become

\vspace*{1.0em}
\begin{tabular}{lccc}
  \toprule
  &FastHenry& BM, (\ref{eq:segmentation method babic})  & MIM, (\ref{eq:segmentation method}) \\
   \midrule
   $M_{\sigma}$, nH &  $-4.65652$ & $-4.686079$& $-4.686079$\\
  \toprule\label{tab:example9}
\end{tabular}

\subsubsection*{ Example 10 }
Using the same arrangement as in Example 9,  the square polygon is inscribed.
The results of calculation are

\vspace*{1.0em}
\begin{tabular}{lccc}
  \toprule
  &FastHenry& BM, (\ref{eq:segmentation method babic})  & MIM, (\ref{eq:segmentation method}) \\
   \midrule
   $M_{\sigma}$, nH &   $-7.07775$ & $-7.094651$& $-7.094651$\\
  \toprule\label{tab:example10}
\end{tabular}

\subsubsection*{ Example 11 }
Using the same arrangement as in Example 9,  the hexagon is inscribed.
The results of calculation are

\vspace*{1.0em}
\begin{tabular}{lccc}
  \toprule
  &FastHenry& BM, (\ref{eq:segmentation method babic})  & MIM, (\ref{eq:segmentation method}) \\
   \midrule
   $M_{\sigma}$, nH &   $-9.0490$ & $-9.0334$& $-9.0334$\\
  \toprule\label{tab:example11}
\end{tabular}

\subsection{ Mutual inductance between two circular filaments}
\label{sec:circle}
In this section, the method of segmentation is applied to the calculation of mutual inductance between two circular filaments arbitrarily orientated in the space with respect to each other. Similar to Sec.\ref{sec:polygon}, the secondary circle is lying  on the $xy$ plane.   The linear misalignment is defined by coordinates of point $C$ corresponding to the centre of the secondary circle, and the angular misalignment is defined by Grover's angles denoted by  $\theta$ and $\eta$ \cite{Poletkin2019}.
The results of the calculation are compared with results obtained by means of  analytical formulas developed by Babic in work \cite{BabicSiroisAkyelEtAl2010} and Poletkin in \cite{Poletkin2019}.
\begin{figure}[!t]
  \centering
  \includegraphics[width=2.9in]{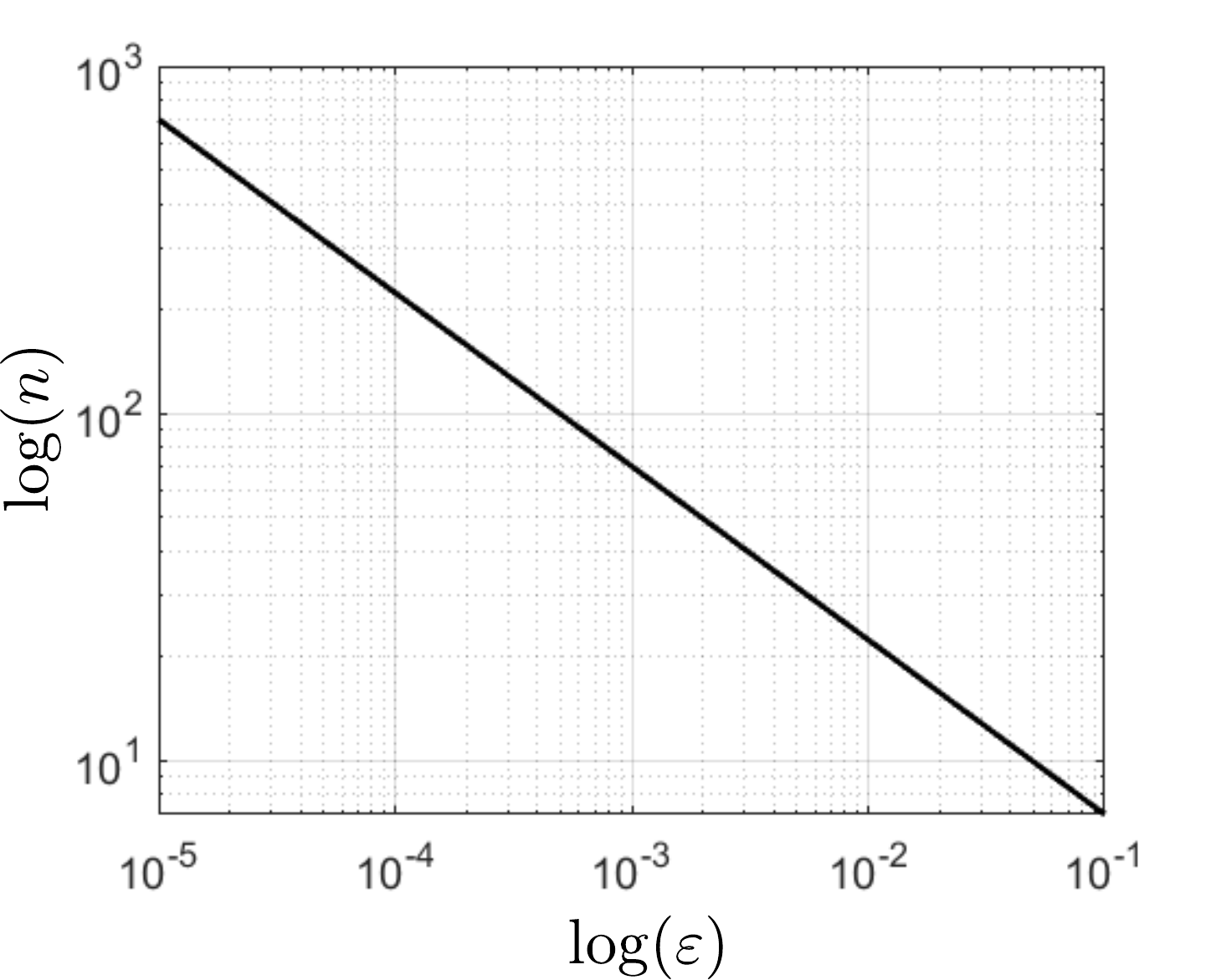}
  \caption{The required number of line segments for given relative error of interpolation of a circle, $n=n(\varepsilon)$.   }\label{fig:error}
\end{figure}

The accuracy of interpolation of a circle by line segments can be measured by the relative error $\varepsilon$, which can be estimated as the deference between the radius of the circle and a length of a line joining a centre of the circle and mid point of a line segment divided by the radius:
\begin{equation}\label{eq:error aprox}
  \varepsilon=1-\cos\frac{\pi}{n}.
\end{equation}
Form 
(\ref{eq:error aprox}) follows the obvious conclusion that the larger the number of line segments the smaller the error. The required number of line segments for the given relative error can be also evaluated 
by (\ref{eq:error aprox}) and presented, for instance, in the logarithmic scale as shown in Fig. \ref{fig:error}. As seen  from Fig. \ref{fig:error}, to interpolate a circle with the relative error, for instance, less than $10^{-3}$, the number of line segments must be more than 100.

For generating input points, equation (\ref{eq:points polygon}) can be used with the only difference  that the number of line segments is being  chosen in according to the desired accuracy of interpolation of a circle, which in turn effects on the accuracy of calculation of mutual inductance.

\begin{figure}[!t]
  \centering
  \includegraphics[width=2.8in]{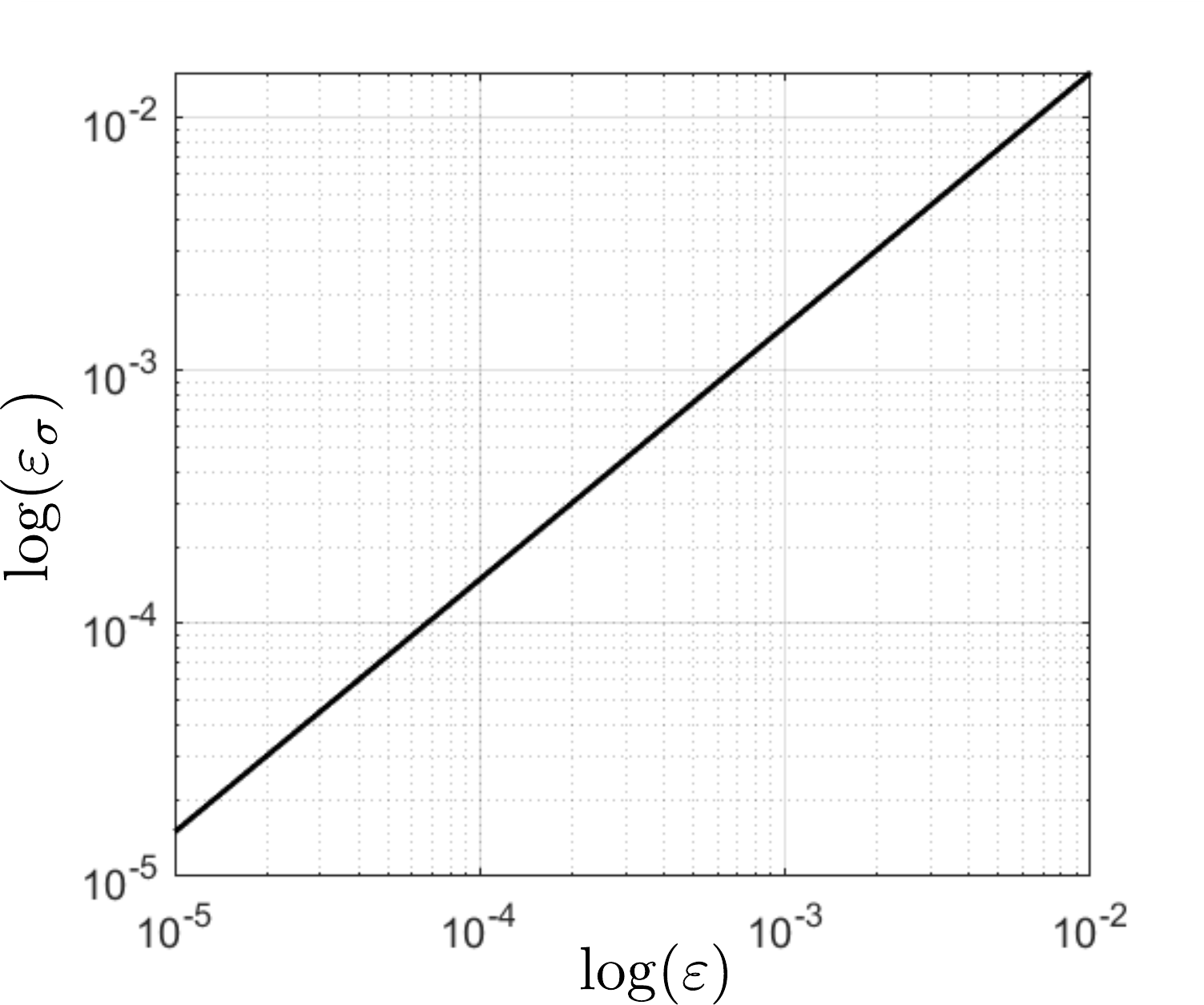}
  \caption{The relative error of calculation of mutual inductance versus  the relative error of interpolation of a circle by line segments.   }\label{fig:error2}
\end{figure}

\begin{figure}[!b]
  \centering
  \includegraphics[width=1.6in]{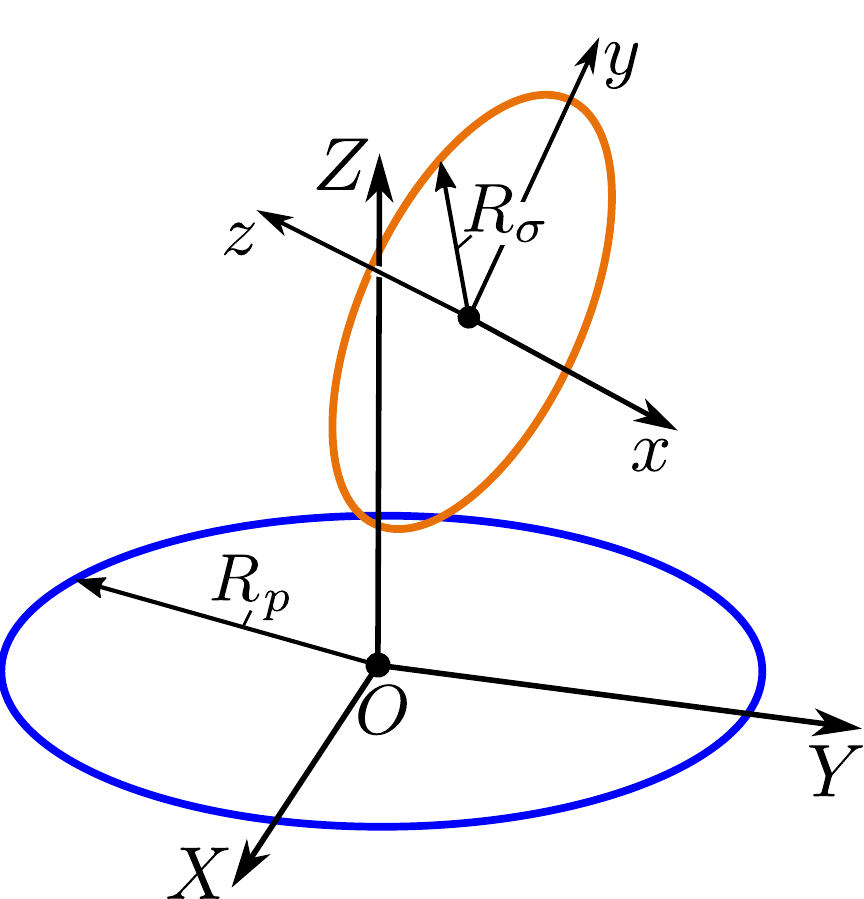}
  \caption{Arrangement of circles in the illustrative example.   }\label{fig:illustrative}
\end{figure}

Indeed, the accuracy of calculation of mutual inductance can be measured also by the relative error, which is defined  as follows
\begin{equation}\label{eq:error}
  \varepsilon_{\sigma}=\frac{M-M_{\sigma}}{M},
\end{equation}
where $M$ is the mutual inductance calculated by Babic's  or Poletkin's formula as mentioned above. Clearly that it is dependent on the number of line segments. To exam this point let us consider the following
geometrical arrangement  taken from Babic's work \cite[page 3597, Example 12]{BabicSiroisAkyelEtAl2010}, as an illustrative example, in which two circles with radii $R_p=$\SI{16.0}{\centi\meter} and $R_{\sigma}=$\SI{10.0}{\centi\meter} and the centre of the secondary  $\sigma$-circle is located at $x_C=0$, $y_C=$\SI{4.3301}{\centi\meter}, $z_C=$\SI{17.5}{\centi\meter} and the angle, $\theta$ of \SI{60.0}{\degree}, but the $\eta$ angle is chosen to be 45.0\si{\degree} as shown in Fig. \ref{fig:illustrative}. Performing calculation for different numbers of line segments,  the results are tabulated as follows

\vspace*{1.0em}
\begin{tabular}{lccc}
  \toprule
  $n$ &$\varepsilon_{\sigma}$& BM, (\ref{eq:segmentation method babic}), nH   & MIM, (\ref{eq:segmentation method}), nH \\
   \midrule
   22 &  $0.015173$ & $15.25271$& $15.25271$\\
    70 &  $0.001507$ & $15.46438$& $15.46438$\\
   222 &  $0.000150$ & $15.48539$& $15.48539$\\
   702 &  $0.000015$ & $15.48748$& $15.48748$\\
  \toprule\label{tab:illustrative example}
\end{tabular}

Using the obtained results shown in the above table, the relative error $\varepsilon_{\sigma}$ as a function of the number $n$ of line segments can be defined, namely, $\varepsilon_{\sigma}=\varepsilon_{\sigma}(n)$. Moreover, both functions   $\varepsilon_{\sigma}=\varepsilon_{\sigma}(n)$ and $n=n(\varepsilon)$ can be joined into one function exhibiting the dependency of the relative error of calculation mutual inductance on  the relative error of interpolation of a circle as shown in Fig. \ref{fig:error2}. Although, Fig. \ref{fig:error2} shows the evaluation of the error for the particular arrangement. However, the result is surprisely consistence and can be used for the preliminary estimation of the error of calculation in other arrangements as shown below. For  further calculation, the number of line segments is set to 200 that, according to  Fig. \ref{fig:error2}, the error, $\varepsilon_{\sigma}$,  is estimated to be  around  $0.00015$.

\begin{figure}[!t]
  \centering
  \includegraphics[width=1.7in]{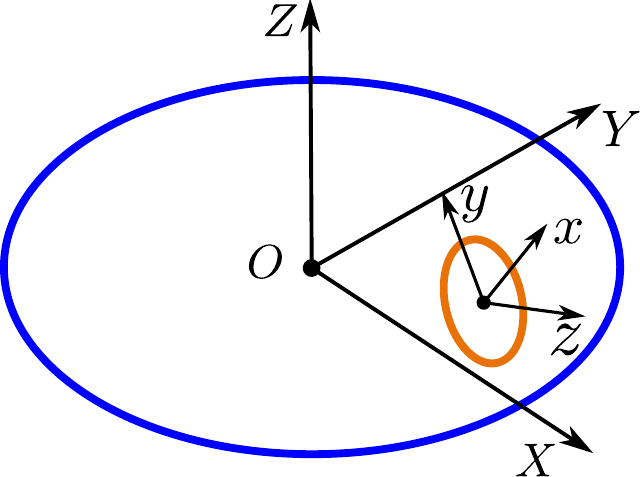}
  \caption{Scheme for Example 12.   }\label{fig:example12}
\end{figure}

\subsubsection*{ Example 12}
The primary and secondary  circles have  radii of $R_p=$\SI{5}{\milli\meter} and $R_s=$\SI{1}{\milli\meter}, respectively. The centre of the secondary circle is located at $x_C=\SI{3}{\milli\meter}$, $y_C=$\SI{1}{\milli\meter}, $z_C=$\SI{0.5}{\milli\meter} and its angular  misalignment are defined by  $\theta=\SI{57.6885}{\degree}$ and $\eta=\SI{108.4349}{\degree}$ as shown in Fig. \ref{fig:example12}. 
The result of calculation is

\vspace*{1.0em}
\begin{tabular}{lccc}
  \toprule
    $n$ &$\varepsilon_{\sigma}$& BM, (\ref{eq:segmentation method babic}), nH   & MIM, (\ref{eq:segmentation method}), nH \\
   \midrule
   200 &  $0.0001566$ & $0.3577388$& $0.3577388$\\
  \toprule\label{tab:example12}
\end{tabular}

\subsubsection*{ Example 13}
The primary and secondary  circles have  radii $R_p=$\SI{40}{\centi\meter} and $R_s=$\SI{5}{\centi\meter}, respectively. The centre of the secondary circle is located at $x_C=\SI{10}{\centi\meter}$, $y_C=$\SI{15}{\centi\meter}, $z_C=$\SI{0.0}{\centi\meter}. The angular misalignment of the secondary filament is defined as follows $\theta=\SI{ 74.4986}{\degree}$ and $\eta=\SI{123.6901}{\degree}$. As it has mentioned above, for calculation the number of line segments of $n=200$ is taken. The result of calculation is

\vspace*{1.0em}
\begin{tabular}{lccc}
  \toprule
    $n$ &$\varepsilon_{\sigma}$& BM, (\ref{eq:segmentation method babic}), nH   & MIM, (\ref{eq:segmentation method}), nH \\
   \midrule
   200 &  $0.0001615$ & $3.848737$& $3.848737$\\
  \toprule\label{tab:example13}
\end{tabular}

\subsubsection*{ Example 14 (Example 11, page 3596 in the Babi\v{c} article \cite{BabicSiroisAkyelEtAl2010})}

Let us consider two circular filaments having radii of $R_p=$\SI{40}{\centi\meter} and $R_s=$\SI{10}{\centi\meter}, which are  mutually perpendicular to each
other that angles of $\eta=$\SI{0}{} and $\theta=$\SI{90.0}{\degree}. The centre of the secondary circle has the following coordinates:  $x_C=0$, $y_C=$\SI{20}{\centi\meter}, and $z_C=$\SI{10}{\centi\meter} as shown in Fig. \ref{fig:example14}.
 The result becomes

\vspace*{1.0em}
\begin{tabular}{lccc}
  \toprule
    $n$ &$\varepsilon_{\sigma}$& BM, (\ref{eq:segmentation method babic}), nH   & MIM, (\ref{eq:segmentation method}), nH \\
   \midrule
   200 &  $0.0001467$ & $-10.72715$& $-10.72715$\\
  \toprule\label{tab:example13}
\end{tabular}
\begin{figure}[!t]
  \centering
  \includegraphics[width=1.7in]{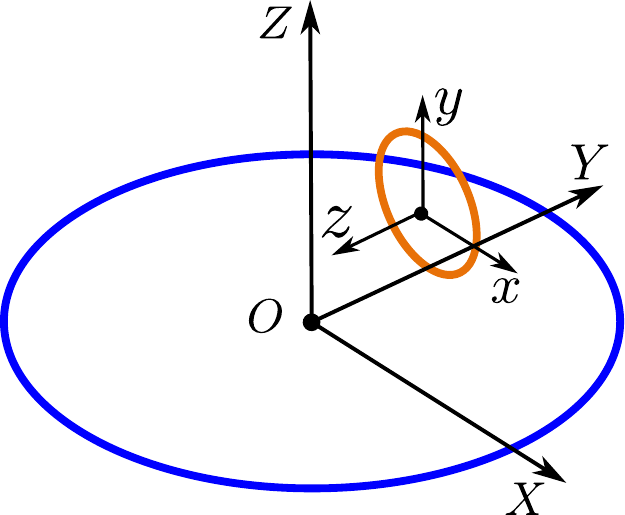}
  \caption{Scheme for Example 14.   }\label{fig:example14}
\end{figure}
\begin{figure}[!t]
  \centering
  \includegraphics[width=3.5in]{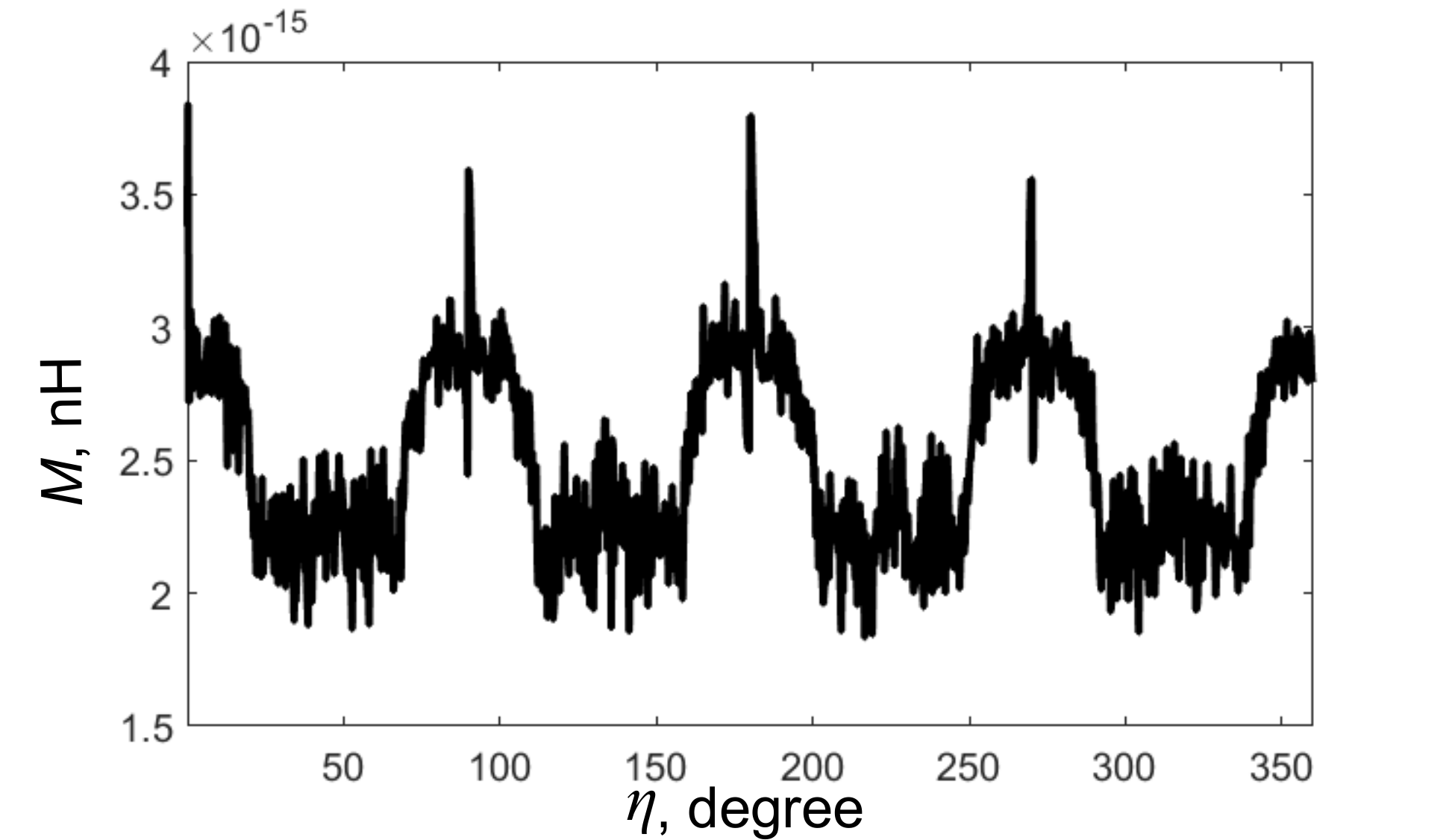}
  \caption{The chaotic distribution of the absolute error against the $\eta$ angle in Example 15.   }\label{fig:chaos1}
\end{figure}

\subsubsection*{ Example 15 (Example 21,  Poletkin's article \cite{Poletkin2019})}
Mutually perpendicular circles (angles of $\theta=$\SI{90.0}{\degree} and $\eta=$\SI{0}{}) having the same radii as in Example 14 are considered, but the centre of the secondary coil is located at the origin $O$. Results of calculation are

\vspace*{1.0em}
\begin{tabular}{lcc}
  \toprule
    $n$ & BM, (\ref{eq:segmentation method babic}), nH   & MIM, (\ref{eq:segmentation method}), nH \\
   \midrule
   200    &$3.27233\times10^{-15}$&$2.91992\times10^{-15}$\\
  \toprule\label{tab:example15}
\end{tabular}

The results show the small error, which has the chaotic nature. Indeed, changing the angle $\eta$ in a range of  $0<\eta\leq360^{o}$ the chaotic distribution of the error  is observed as shown in Fig. \ref{fig:chaos1}. Analysis of the figure depicts that the error does not exceed of  \SI{4e-15}{\nano\henry}. Worth noting that varying the number of line segments, $n$, alters the error distribution, but keeps its value below    the observed maximum.

\subsubsection*{ Example 16 (Example 22,  Poletkin's article \cite{Poletkin2019}) }
Now we again consider mutually perpendicular circles having the same radii as in Example 14, but in this case the centre of the secondary circle occupies a position on the $XOY$-surface with the following coordinates   $x_C=y_C=$\SI{10}{\centi\meter} and $z_C=0$. Hence, we have

\vspace*{1.0em}
\begin{tabular}{lcc}
  \toprule
   $n$ & BM, (\ref{eq:segmentation method babic}), nH   & MIM, (\ref{eq:segmentation method}), nH \\
   \midrule
   200    &$-6.23859\times10^{-15}$&$-8.05774\times10^{-15}$\\
  \toprule\label{tab:example15}
\end{tabular}

\begin{figure}[!t]
  \centering
  \includegraphics[width=3.5in]{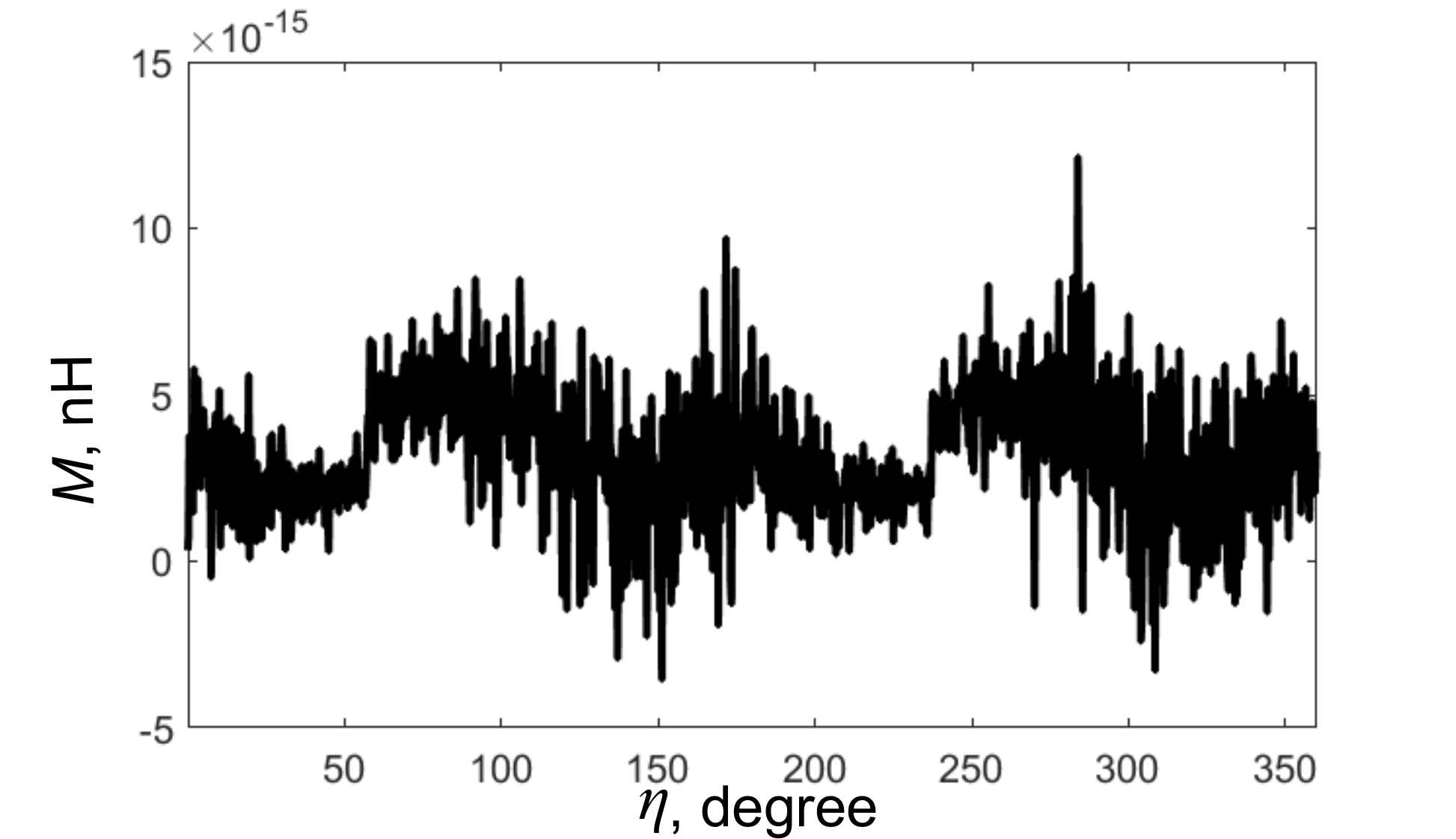}
  \caption{The chaotic distribution of the absolute error against the $\eta$ angle in Example 16.   }\label{fig:chaos2}
\end{figure}
Then, again let us rotate the $\eta$ angle in a range $0<\eta\leq360^{o}$,  the calculation of mutual inductance demonstrates the chaotic distribution of the small error of calculation, which does not exceeded   of  \SI{15e-15}{\nano\henry} as shown in Fig. \ref{fig:chaos2}. The error is continually distributed without interruption of its continuity  at $\eta= $\SI{90}{\degree} and \SI{270}{\degree} in comparing with the application of Babi{c}'s and Grover's formula  (please, see Example 22,  in article \cite{Poletkin2019}).
Thus, the segmentation method for the calculation of mutual inductance between two circular filaments arbitrarily orientated in the space with respect to each other does not suffer from the singularity case, when two circles are mutually perpendicular.

\subsection{ Mutual inductance between circle and  a filament of arbitrary shape}
\label{sec:arb}

Using the proposed methodology based on   (\ref{eq:segmentation method}) and (\ref{eq:segmentation method babic}), the segmentation method is applied to the calculation of the mutual inductance between the primary circle and filaments arbitrary positioning in the space and having different shapes, which can be defined, for instance, by parametric equations of  special curves to generate the input set of points (\ref{eq:p}). In particular, the following curves such as  circular and elliptic arcs, ellipse, spiral, helices and conical helices are considered in the proceeding section.

\begin{figure}[!t]
  \centering
  \includegraphics[width=3.5in]{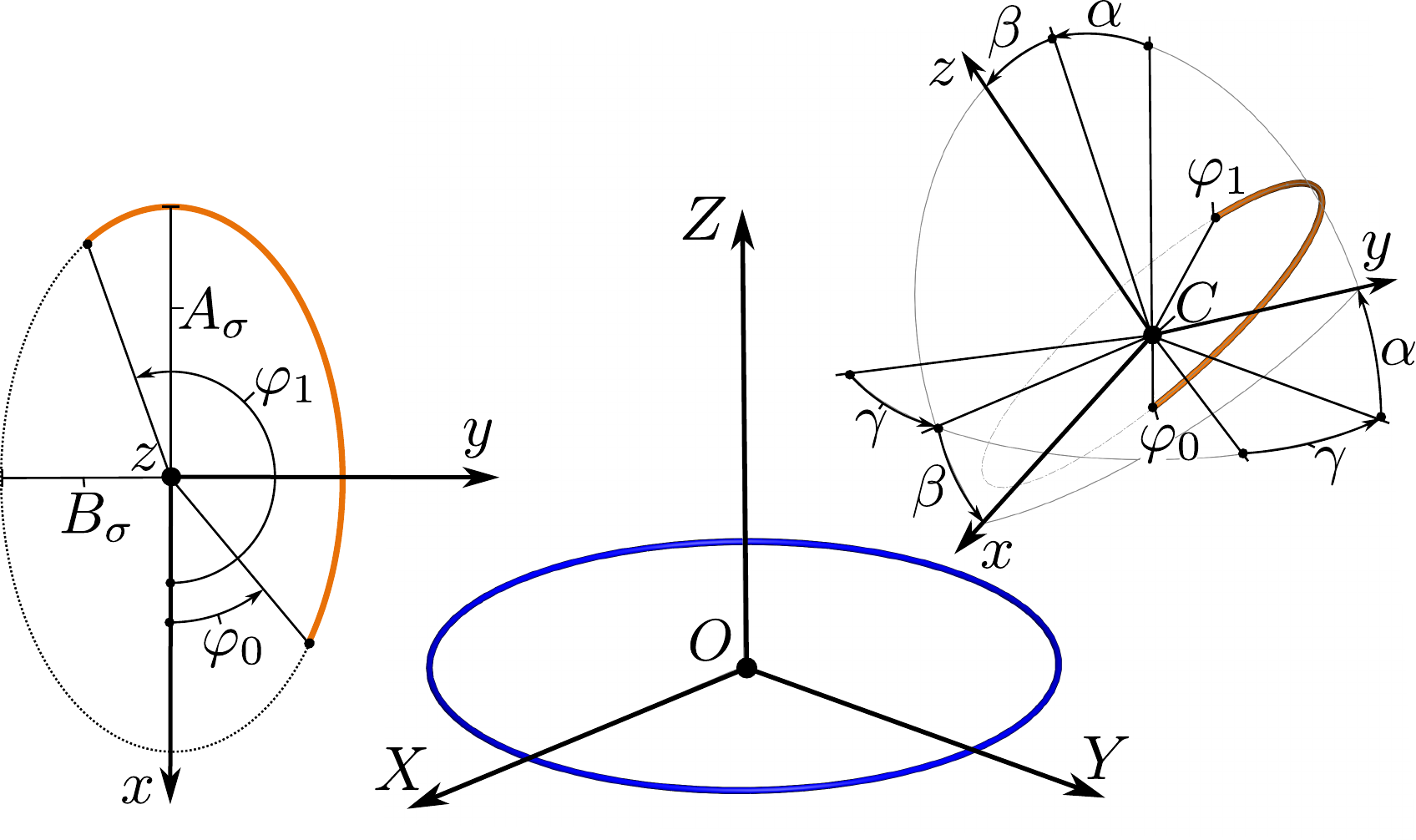}
  \caption{ Elliptic arc on the $xy$ plane is determined as a part of ellipse having the semi-major axis $A_{\sigma}$ and semi-minor $B_{\sigma}$  within the finite interval $\left[\varphi_0,\varphi_1\right]$ (the left side figure);  The position of elliptic arc arbitrary orientated in the space with respect to the primary circle (the right side figure).   }\label{fig:arc}
\end{figure}

Beginning with the calculation of mutual inductance between the circle and elliptic arc arbitrary positioning in the space, the secondary filament can be defined as a part of ellipse lying on the $xy$ plane. The ellipse is characterized by its semi-major axis $A_{\sigma}$ directed along the $x$-axis and  semi-minor $B_{\sigma}$ along the $y$-axis. The parameter $\ell$ is defined within the finite interval $\left[\varphi_0,\varphi_1\right]$ as shown in Fig. \ref{fig:arc}. The linear misalignment of the arc is defined by coordinates of point $C$ corresponding to the centre of the ellipse, while the angular misalignment can be determined by Euler's angles 
denoted by $\alpha$, $\beta$ and $\gamma$. The following relationship between Grover's  and Euler's angles   is true \cite{Poletkin2019}:
\begin{equation}\label{eq:angles}
  \left\{\begin{array}{l}
   \sin\beta=\sin\eta\sin\theta;\\
   \cos\beta\sin\alpha=\cos\eta\sin\theta.
  \end{array}\right.
\end{equation}

According to Fig. \ref{fig:arc},  the input points can be calculated as follows
\begin{equation}\label{eq:points ellipces}
 \underline{p}_i(h_i)=\left[\begin{array}{c} x_C \\ y_C \\  z_C \end{array}\right]+ \underline{\Lambda}_{\gamma}^T\underline{\Lambda}_{\alpha}^T \underline{\Lambda}_{\beta}^T\left[\begin{array}{c} A_{\sigma}\cos h_i \\ B_{\sigma}\sin h_i \\  0\end{array}\right], \; i=0\ldots n,
\end{equation}
where
\begin{equation}\label{eq:cosine matrix Euler}
\begin{array}{c}
  \underline{\Lambda}_{\gamma}=\left[\begin{array}{ccc} \cos\gamma & \sin\gamma & 0\\ -\sin\gamma & \cos\gamma & 0\\ 0 & 0 & 1 \end{array}\right], \;
 \underline{\Lambda}_{\alpha}=\left[\begin{array}{ccc} 1 & 0 & 0\\ 0 & \cos\alpha & \sin\alpha\\ 0 & -\sin\alpha & \cos\alpha \end{array}\right],\\
  \underline{\Lambda}_{\beta}=\left[\begin{array}{ccc} \cos\beta & 0 & -\sin\beta\\ 0 & 1 & 0\\ \sin\beta & 0 & \cos\beta \end{array}\right].
\end{array}
\end{equation}

\subsubsection*{ Example 17}
The primary circle has a radius of $R_p=$\SI{1}{\meter}, while the elliptic arc is defined by the ellipse having
the semi-major axis $A_{\sigma}=$ \SI{1}{\meter} and  semi-minor one $B_{\sigma}=$\SI{0.5}{\meter} and within
the finite interval $\left[\varphi_0=\SI{10}{\degree},\varphi_1=\SI{110}{\degree}\right]$.
  The centre of the ellipse is located at $x_C=\SI{10.0}{\centi\meter}$, $y_C=$\SI{10.0}{\centi\meter}, $z_C=$\SI{10.0}{\centi\meter}. The angular misalignment of the secondary filament  is defined by  $\alpha=\SI{ 20.0}{\degree}$, $\beta=\SI{20.0}{\degree}$, but  $\gamma$ is changed in a range from 0 to \SI{325.0}{\degree}. The number of line segments of $n=200$ is taken. The results of calculation are

\vspace*{1.0em}
\begin{tabular}{lccc}
  \toprule
  $\gamma$&FastHenry, nH & BM,(\ref{eq:segmentation method babic}), nH   & MIM, (\ref{eq:segmentation method}), nH\\
   \midrule
 \SI{0.0}{\degree}& $368.066$    &$368.191$&$368.191$\\
  \SI{35.0}{\degree}& $330.527$    &$329.896$&$329.896$\\
  \SI{100.0}{\degree}& $243.600$    &$242.784$&$242.784$\\
  \SI{180.0}{\degree}& $214.997$    &$214.6251$&$214.6251$\\
  \SI{250.0}{\degree}& $288.439$    &$289.2692$&$289.2692$\\
  \SI{300.0}{\degree}& $354.949$    &$356.677$&$356.677$\\
  \SI{325.0}{\degree}& $373.121$    &$375.319$&$375.319$\\
  \toprule\label{tab:example17}
\end{tabular}

It is clear that an elliptic arc can be transformed into the other curves   such as a circular one ($A_{\sigma}=B_{\sigma}$, $\left[\varphi_0\geq0,\varphi_1<2\pi\right]$), an ellipse ($A_{\sigma}\neq B_{\sigma}$, $\left[\varphi_0=0,\varphi_1=2\pi\right]$) and a circle ($A_{\sigma}=B_{\sigma}$, $\left[\varphi_0=0,\varphi_1=2\pi\right]$).

\begin{figure}[!t]
  \centering
  \includegraphics[width=2.0in]{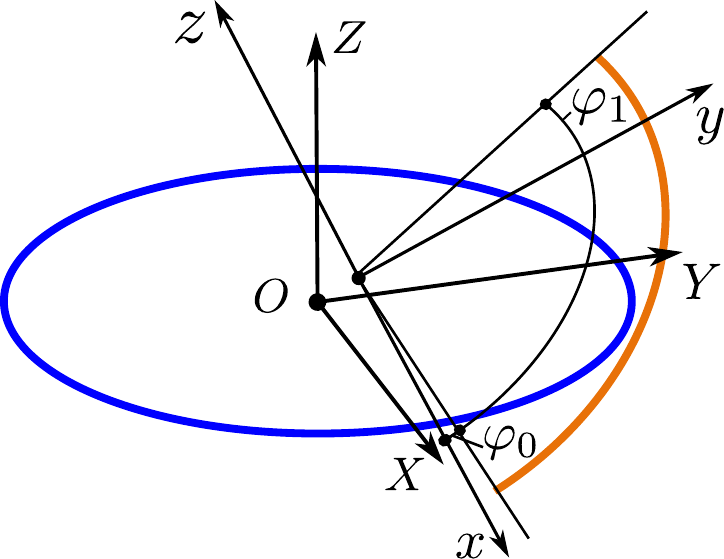}
  \caption{ The scheme for Example 18 (circular arc): $\alpha=\SI{ 20.0}{\degree}$, $\beta=\SI{20.0}{\degree}$ and $\gamma=\SI{0.0}{\degree}$ .   }\label{fig:example18}
\end{figure}
\subsubsection*{ Example 18 (circular arc)}
The primary circle has a radius of $R_p=$\SI{1}{\meter}, while the circular arc is defined by  $A_{\sigma}=B_{\sigma}=$\SI{1}{\meter}  within
the finite interval $\left[\varphi_0=\SI{10}{\degree},\varphi_1=\SI{110}{\degree}\right]$.
  The centre of the secondary circle is located at $x_C=\SI{10.0}{\centi\meter}$, $y_C=$\SI{10.0}{\centi\meter}, $z_C=$\SI{10.0}{\centi\meter}. The angular misalignment of the secondary filament  is defined as follows $\alpha=\SI{ 20.0}{\degree}$, $\beta=\SI{20.0}{\degree}$ but  $\gamma$ is changed in a range from 0 to \SI{325.0}{\degree} as shown in Fig. \ref{fig:example18}. The number of line segments of $n=200$ is taken. The results of calculation are

\vspace*{1.0em}
\begin{tabular}{lccc}
  \toprule
   $\gamma$&FastHenry, nH & BM, (\ref{eq:segmentation method babic}), nH   & MIM, (\ref{eq:segmentation method}), nH\\
   \midrule
 \SI{0.0}{\degree}& $452.190$    &$452.632$&$452.632$\\
  \SI{35.0}{\degree}& $463.948$    &$463.5728$&$463.5728$\\
  \SI{100.0}{\degree}& $577.401$    &$577.2027$&$577.2027$\\
  \SI{180.0}{\degree}& $461.735$    &$461.3058$&$461.3058$\\
  \SI{250.0}{\degree}& $522.928$    &$523.709$&$523.709$\\
  \SI{300.0}{\degree}& $517.435$    &$519.920$&$519.920$\\
  \SI{325.0}{\degree}& $474.681$    &$477.143$&$477.143$\\
  \toprule\label{tab:example17}
\end{tabular}

\subsubsection*{ Example 19 (ellipse)}
The primary circle has a radius of $R_p=$\SI{1}{\meter}, while the ellipse is defined by  $A_{\sigma}=$ \SI{1.0}{\meter} $B_{\sigma}=$\SI{0.5}{\meter}.
  The centre of the ellipse is located at $x_C=\SI{10.0}{\centi\meter}$, $y_C=$\SI{10.0}{\centi\meter}, $z_C=$\SI{10.0}{\centi\meter}. The angular misalignment of the secondary filament  is defined by $\alpha=\SI{ 20.0}{\degree}$, $\beta=\SI{20.0}{\degree}$ and  $\gamma=\SI{0.0}{\degree}$ as shown in Fig. \ref{fig:example19}. The number of line segments of $n=200$ is taken. The result of calculation is

\begin{figure}[!t]
  \centering
  \includegraphics[width=2.1in]{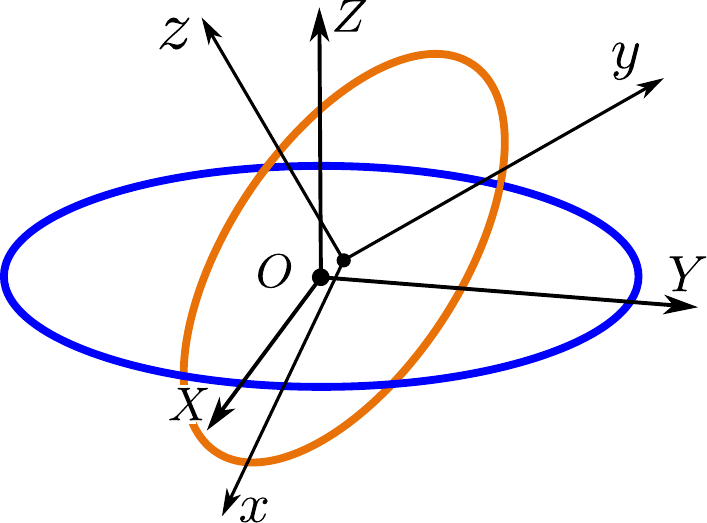}
  \caption{ The scheme for Example 19 (ellipse): $\alpha=\SI{ 20.0}{\degree}$, $\beta=\SI{20.0}{\degree}$ and $\gamma=\SI{0.0}{\degree}$ .   }\label{fig:example19}
\end{figure}
\vspace*{1.0em}
\begin{tabular}{ccc}
  \toprule
  FastHenry, nH & BM, (\ref{eq:segmentation method babic}), nH   & MIM, (\ref{eq:segmentation method}), nH\\
   \midrule
  $904.848$    &$905.9695$&$905.9695$\\
  \toprule\label{tab:example19}
\end{tabular}

\begin{figure}[!t]
  \centering
  \includegraphics[width=3.3in]{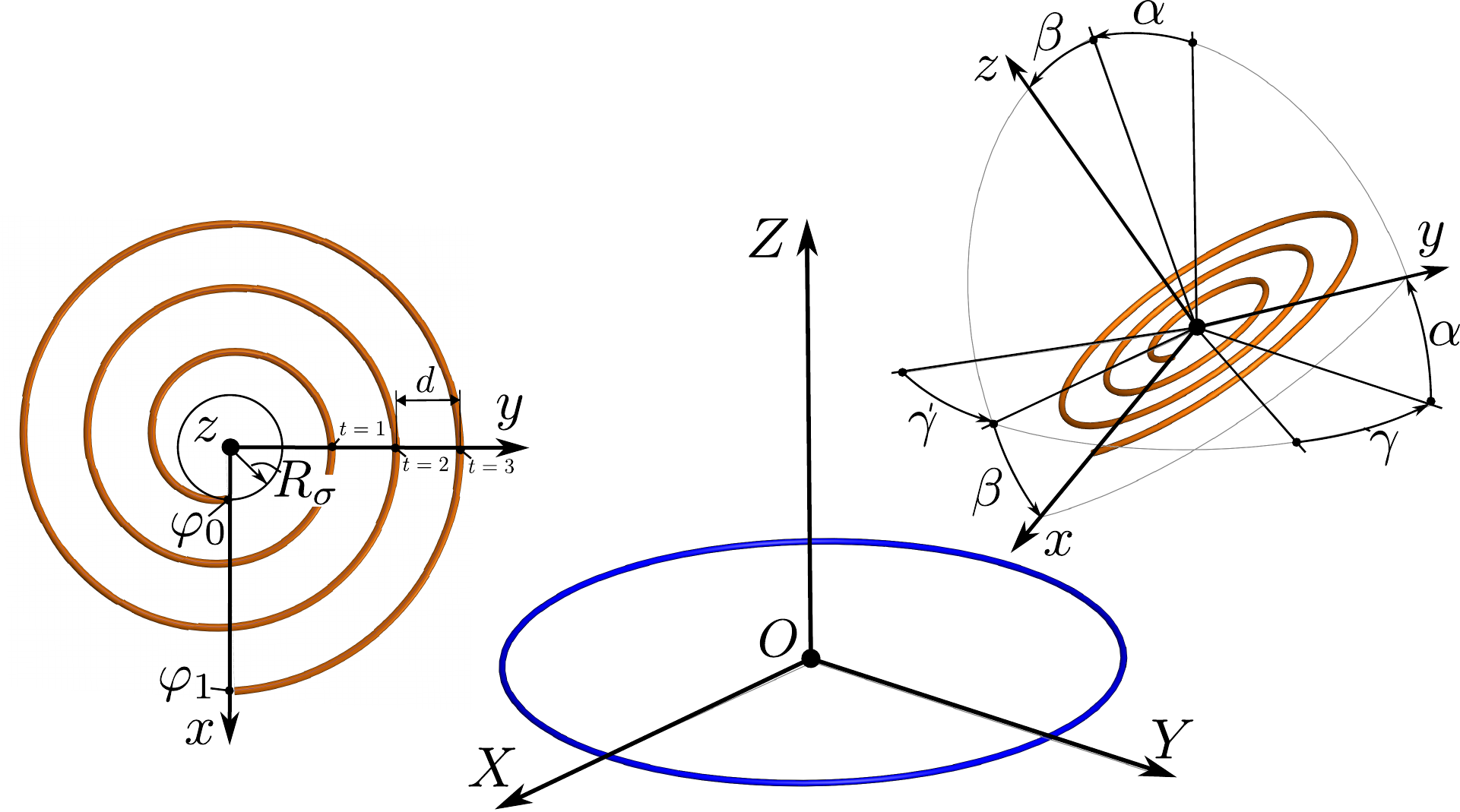}
  \caption{ Spiral on the $xy$ plane is determined by the inner radius of $R_{\sigma}$, the distance between turns $d$ and number of turns $t$ (the left side figure);  The position of the spiral arbitrary orientated in the space with respect to the primary circle (the right side figure).   }\label{fig:spiral}
\end{figure}
For the case of  the calculation of mutual inductance  between the circle and a spiral, the secondary filament is lying on the $xy$ plane and  its geometry  can be characterized by the inner radius $R_{\sigma}$, the number of turns $t$ and the distance between the turns $d$ as shown on Fig. \ref{fig:spiral}. The boundaries of the finite interval of the parameter $\ell$ are estimated as follows
  \begin{equation}\label{eq:boundaries}
    {\displaystyle \varphi_0=2\pi R_{\sigma}/d,\; \varphi_1=\varphi_0+2\pi t.} \\
  \end{equation}
Similar to the elliptic arc,  the linear misalignment of the spiral is defined by coordinates of point $C$, and the angular misalignment is defined by Euler's angles as shown in Fig. \ref{fig:spiral}. The input points are generated by the following equation:
\begin{equation}\label{eq:points spiral}
 \underline{p}_i(h_i)=\left[\begin{array}{c} x_C \\ y_C \\  z_C \end{array}\right]+  \underline{\Lambda}_{\gamma}^T\underline{\Lambda}_{\alpha}^T \underline{\Lambda}_{\beta}^T\left[\begin{array}{c} {\displaystyle\frac{ d h_i}{2\pi}}\cos h_i \\   {\displaystyle\frac{ d h_i}{2\pi}}\sin h_i \\  0\end{array}\right], \; i=0\ldots n,
\end{equation}
where  cosine matrices  are determined by Eq. (\ref{eq:cosine matrix Euler}).

\subsubsection*{ Example 20 (spiral)}
The primary circle has a radius of $R_p=$\SI{0.5}{\meter}. The spiral is defined by the following parameters: the inner radius $R_{\sigma}$ is \SI{0.05}{\meter}, the distance $d$ is \SI{0.04}{\meter} and the number of turns $t$ is $9$.
  The centre $C$ is located at $x_C=\SI{0.6}{\meter}$, $y_C=$\SI{0.1}{\meter}, and $z_C=$\SI{0.7}{\meter}. The angular misalignment  is defined as follows $\alpha=\SI{ 45.0}{\degree}$, $\beta=\SI{-45.0}{\degree}$, 
  but  $\gamma$ is changed in a range from $0$ to \SI{350.0}{\degree}.
  The number of line segments $n$ is chosen to be  $900$ (It is recommended to use at least 100 line segments per turn to have a reasonable accuracy). The results of calculation are

\vspace*{1.0em}
\begin{tabular}{lccc}
  \toprule
   $\gamma$&FastHenry, nH & BM, (\ref{eq:segmentation method babic}), nH   & MIM, (\ref{eq:segmentation method}), nH\\
   \midrule
 \SI{0.0}{\degree}& $-72.7737$    &$-73.645$&$-73.645$\\
  \SI{35.0}{\degree}& $8.58253$    &$8.73071$&$8.73071$\\
  \SI{100.0}{\degree}& $179.826$    &$180.625$&$180.625$\\
  \SI{180.0}{\degree}& $217.839$    &$217.859$&$217.859$\\
  \SI{250.0}{\degree}& $74.2666$    &$73.3870$&$73.3870$\\
  \SI{300.0}{\degree}& $-63.6821$    &$-64.2813$&$-64.2813$\\
  \SI{350.0}{\degree}& $-86.1984$    &$-87.1184$&$-87.1184$\\
  \toprule\label{tab:example20}
\end{tabular}

\begin{figure}[!t]
  \centering
  \includegraphics[width=3.5in]{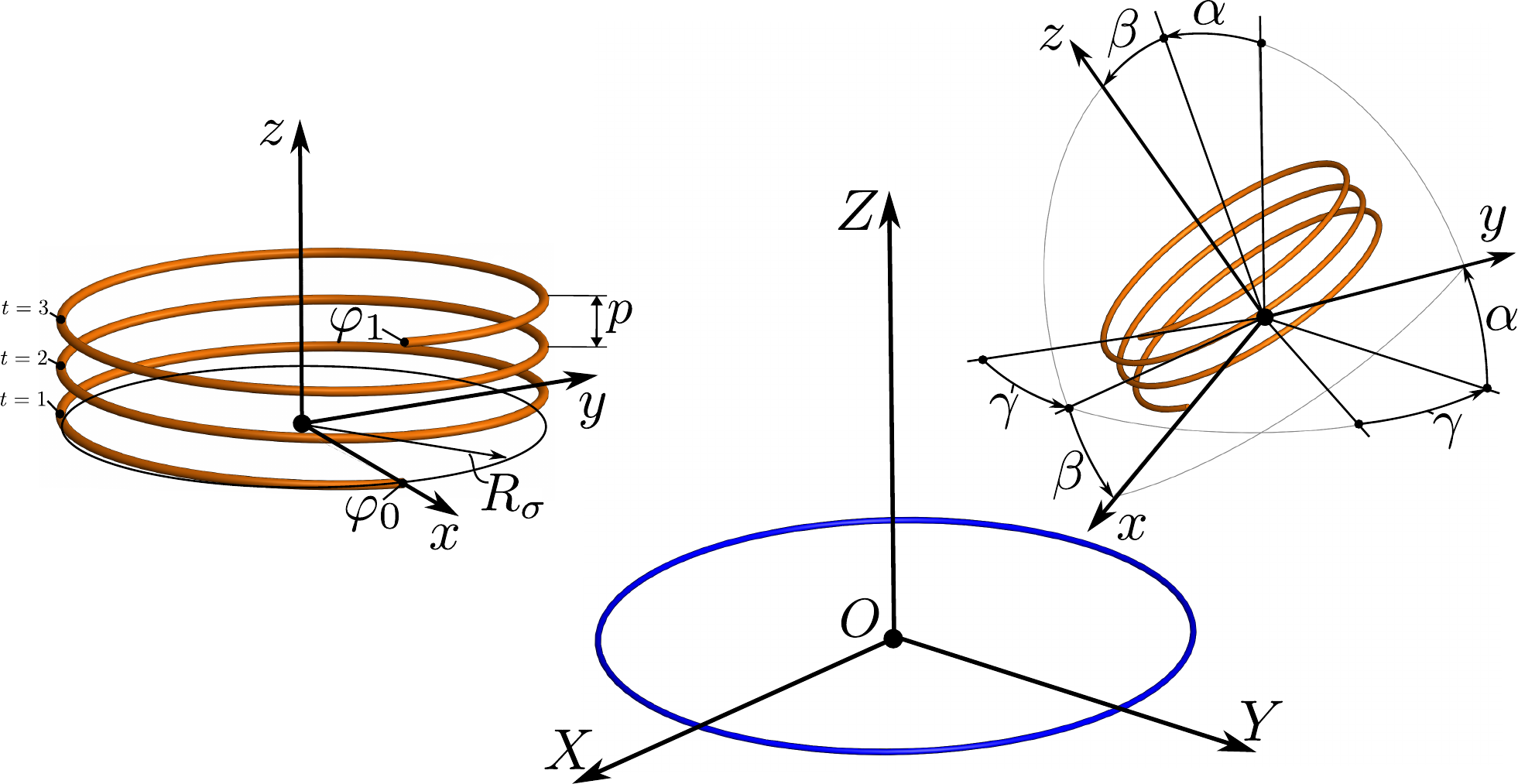}
  \caption{The helices is determined by its radius  $R_{\sigma}$, the pitch $p$ and the number of turns $t$ (the left side figure);  The position of the helices arbitrary orientated in the space with respect to the primary circle (the right side figure).   }\label{fig:helices}
\end{figure}
Further,   the calculation of mutual inductance  between the circle and a helices is considered. The geometry of  helices  can be defined on the $xyz$ CF by its radius $R_{\sigma}$, the number of turns $t$ and the pitch $p$ as shown in Fig. \ref{fig:helices}. The boundaries of the finite interval of the parameter $\ell$ are
  \begin{equation}\label{eq:boundaries}
    {\displaystyle \varphi_0=0,\; \varphi_1=2\pi t.} \\
  \end{equation}
Similar to previous cases,  the linear misalignment of the helices is defined by coordinates of origin $C$ of the $xyz$ CF, and the angular misalignment is defined by Euler's angles as shown in Fig. \ref{fig:helices}. The following equation can be used to generate the input points:
 \begin{equation}\label{eq:points helices}
 \underline{p}_i(h_i)=\left[\begin{array}{c} x_C \\ y_C \\  z_C \end{array}\right]+  \underline{\Lambda}_{\gamma}^T\underline{\Lambda}_{\alpha}^T \underline{\Lambda}_{\beta}^T\left[\begin{array}{c}  R_{\sigma}\cos h_i \\   {\displaystyle -R_{\sigma}}\sin h_i \\  {\displaystyle \frac{ph_i}{2\pi} }\end{array}\right], \; i=0\ldots n.
\end{equation}

 \subsubsection*{{ Example 21 (helices)}}
The primary circle has  radius of $R_p=$\SI{0.9}{\meter}.
The helices is defined by the following parameters: the  radius $R_{\sigma}$ is \SI{0.6}{\meter}, the pitch $p$ is \SI{0.05}{\meter} and the number of turns $t$ is $4$.
 The origin $C$ is located at $x_C=\SI{0.3}{\meter}$, $y_C=$\SI{0.2}{\meter}, $z_C=$\SI{0.5}{\meter}. The  angular misalignment is characterized by $\alpha=\SI{54.7356}{\degree}$, $\beta=\SI{0.0}{\degree}$ and $\gamma$ is taken from a range from $0$ to \SI{335.0}{\degree}. The number of line segments $n$ is chosen to be  $400$.  The results of calculation are

\vspace*{1.0em}
\begin{tabular}{lccc}
  \toprule
     $\gamma$&FastHenry, nH & BM, (\ref{eq:segmentation method babic}), nH   & MIM, (\ref{eq:segmentation method}), nH\\
   \midrule
 \SI{0.0}{\degree}& $-963.374$    &$-965.106$&$-965.106$\\
  \SI{35.0}{\degree}& $-1242.69$    &$-1245.91$&$-1245.91$\\
  \SI{135.0}{\degree}& $-1892.12$    &$-1893.43$&$-1893.43$\\
  \SI{180.0}{\degree}& $-1537.51$    &$-1537.38$&$-1537.38$\\
  \SI{235.0}{\degree}& $-1025.08$    &$-1024.81$&$-1024.81$\\
  \SI{300.0}{\degree}& $-771.485$    &$-771.878$&$-771.878$\\
  \SI{335.0}{\degree}& $-835.499$    &$-836.041$&$-836.041$\\
  \toprule\label{tab:example21}
\end{tabular}

\begin{figure}[!t]
  \centering
  \includegraphics[width=3.4in]{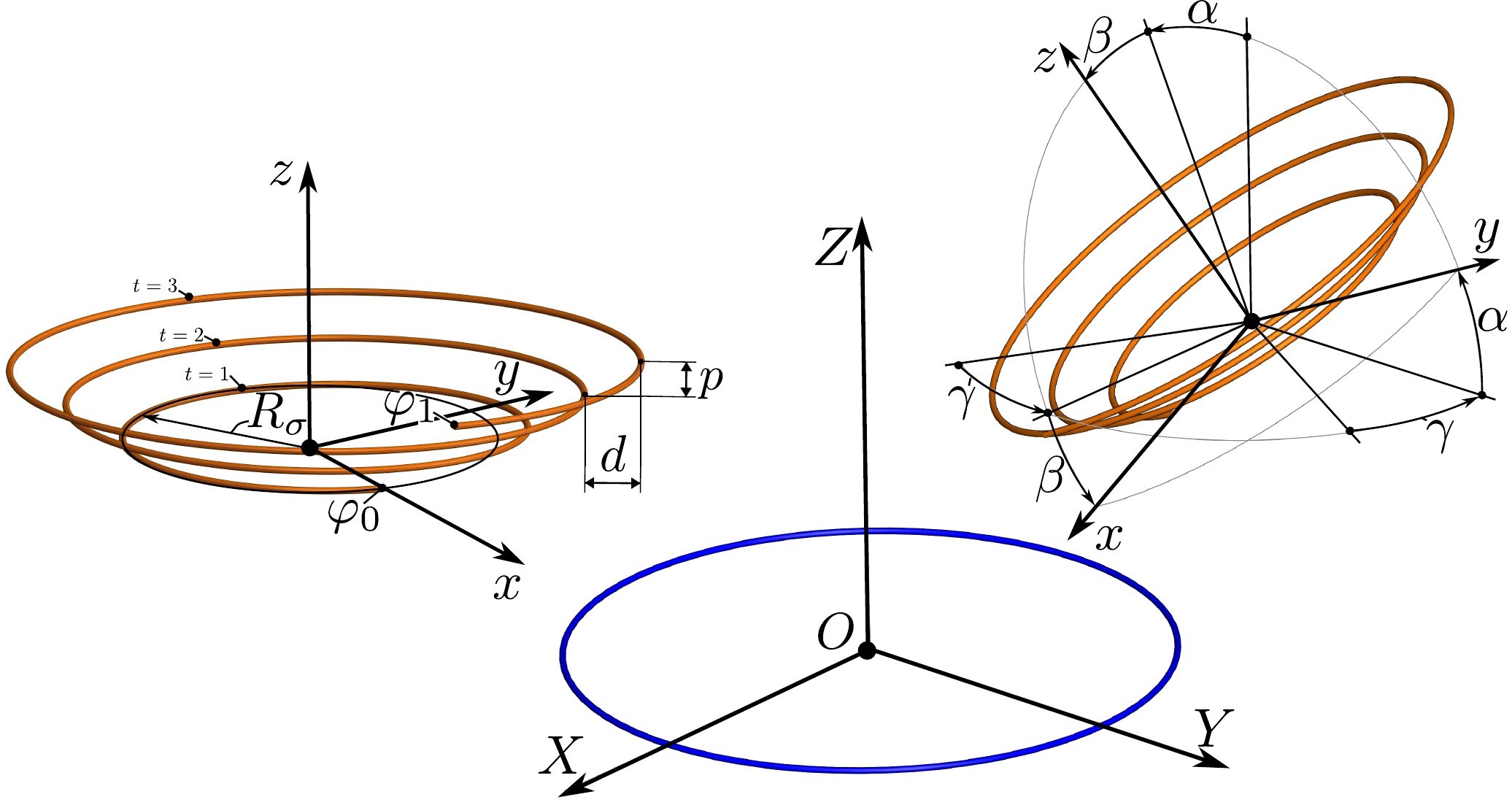}
  \caption{The conical helices is determined by its radius  $R_{\sigma}$, the pitch $p$,  the number of turns $t$ and the distance between the turns $d$ (the left side figure);  The position of the conical helices arbitrary orientated in the space with respect to the primary circle (the right side figure).   }\label{fig:chelices}
\end{figure}
Finally,   the calculation of mutual inductance  between the circle and a conical helices is carried out. The geometry of  helices  can be defined on the $xyz$ CF by its radius $R_{\sigma}$, the number of turns $t$, the pitch $p$ and the distance between turns $d$ as shown in Fig. \ref{fig:chelices}. The boundaries of the finite interval of the parameter $\ell$ are
  \begin{equation}\label{eq:boundaries}
    {\displaystyle \varphi_0=2\pi R_{\sigma}/d,\; \varphi_1=\varphi_0+2\pi t.} \\
  \end{equation}
The linear misalignment of the conical helices is defined by coordinates of origin $C$ of the $xyz$ CF, and the angular misalignment is defined by Euler's angles as shown in Fig. \ref{fig:chelices}. The following equation  generates the input points:
 \begin{equation}\label{eq:points chelices}
 \underline{p}_i(h_i)=\left[\begin{array}{c} x_C \\ y_C \\  z_C \end{array}\right]+  \underline{\Lambda}_{\gamma}^T\underline{\Lambda}_{\alpha}^T \underline{\Lambda}_{\beta}^T\left[\begin{array}{c}  R_{\sigma}h_i\cos h_i \\   {\displaystyle -R_{\sigma}}h_i\sin h_i \\  {\displaystyle \frac{p(h_i-\varphi_0)}{2\pi} }\end{array}\right], \; i=0\ldots n.
\end{equation}

 \subsubsection*{{ Example 22 (conical helices)}}
The primary circle has  radius of $R_p=$\SI{0.9}{\meter}.
The conical helices is defined by the following parameters: the  radius $R_{\sigma}$ is \SI{0.0}{\meter}, the pitch $p$ is \SI{0.1}{\meter},  the number of turns $t$ is $4$ and the distance $d$ is \SI{0.04}{\meter}.
 The origin $C$ is located at $x_C=\SI{0.3}{\meter}$, $y_C=$\SI{0.2}{\meter}, $z_C=$\SI{0.5}{\meter}. The  angular misalignment is characterized by $\alpha=\SI{135}{\degree}$, $\beta=\SI{0.0}{\degree}$ and $\gamma$ is taken from  $0$ to \SI{335.0}{\degree} as shown in Fig. \ref{fig:example22}. The number of line segments $n$ is chosen to be  $400$.  The results of calculation are

\begin{figure}[!t]
  \centering
  \includegraphics[width=1.8in]{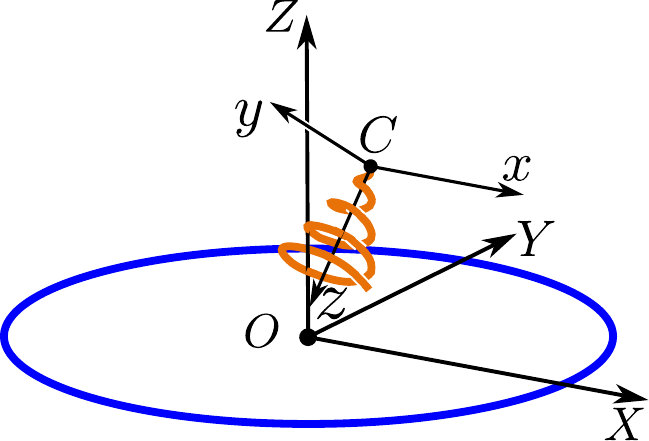}
  \caption{ The scheme for Example 22 (conical helices): $\alpha=\SI{ 135.0}{\degree}$, $\beta=\SI{0.0}{\degree}$ and $\gamma=\SI{0.0}{\degree}$ .   }\label{fig:example22}
\end{figure}
\vspace*{1.0em}
\begin{tabular}{lccc}
  \toprule
     $\gamma$&FastHenry, nH & BM, (\ref{eq:segmentation method babic}), nH   & MIM, (\ref{eq:segmentation method}), nH\\
   \midrule
 \SI{0.0}{\degree}& $27.2802$    &$27.2225$&$27.2225$\\
  \SI{35.0}{\degree}& $29.1615$    &$28.8732$&$28.8732$\\
  \SI{135.0}{\degree}& $68.2821$    &$68.401$&$68.401$\\
  \SI{180.0}{\degree}& $84.7305$    &$85.15068$&$85.15068$\\
  \SI{235.0}{\degree}& $79.8669$    &$79.6694$&$79.6694$\\
  \SI{300.0}{\degree}& $47.0447$    &$46.7936$&$46.7936$\\
  \SI{335.0}{\degree}& $32.3496$    &$32.1987$&$32.1987$\\
  \toprule\label{tab:example22}
\end{tabular}

\section{Conclusion}

In this article, two novel formulas for calculation of mutual inductance between a circular filament and line segment arbitrarily positioning in the space were derived by means of mutual inductance method and Babic's method, respectively. These two formulas were expressed via the integral relationships, whose kernels  include the elliptic functions of the first and second kinds. The two formulas are successfully validated to each other mutually and numerically via the \textit{FastHenry} software.

Using the fact that any curve can be interpolated with a desired accuracy by a finite number of line segments, a segmentation method for calculation of the mutual inductance between the primary
circle and a filament having an arbitrary shape is proposed
and successfully developed based on the two derived formulas.
It is shown that  for calculation of mutual inductance between the primary circle and the secondary filament of an arbitrarily shape  by means of segmentation method based on   (\ref{eq:segmentation method}) and (\ref{eq:segmentation method babic}),  the set of points belonging to the  secondary filament as the input data is only needed.

The proposed segmentation method 
is successfully applied to the calculation of the mutual inductance between the primary circle and filaments arbitrary positioning in the space and having different shapes such as  circular and elliptic arcs, ellipse, spiral, helices and conical helices. All results of calculation are in a  good agreement with the reference examples and successfully validated by   the \textit{FastHenry} software.

Considering a limit case, when the radius of the primary circle is trending to zero, the developed methodology can be used for calculation of the magnetic flux density and its gradient generated by a current-carrying arbitrarily shaped filament. As a result, the force and stiffness between two arbitrarily shaped current-carrying filaments can be calculated by using developed methodology. This matter can be considered as our future work.


%

%
%

\section*{Acknowledgment}

 S.S.K., E.R.M., and K.V.P. acknowledge with thanks the support from German Research Foundation (Grant KO 1883/37-1) under the priority programme SPP 2206.

\ifCLASSOPTIONcaptionsoff
  \newpage
\fi



\bibliographystyle{IEEEtran}
\bibliography{References}
\end{document}